\pdfoutput=1
\documentclass[twocolumn,superscriptaddress,prb]{revtex4}       
\usepackage{color,graphicx}                               
\usepackage{amssymb}                                      
\usepackage{lineno}                                            

\begin{document}
\title{The magnetic and structural properties near the Lifshitz point in Fe$_{1+x}$Te }

\author{E. E. Rodriguez}
\affiliation{Department of Chemistry of Biochemistry, University of Maryland, College Park, MD, 20742, U.S.A.}
\author{D. A. Sokolov}
\affiliation{School of Physics and Astronomy, University of Edinburgh, Edinburgh EH9 3JZ, U.K.}
\author{C. Stock}
\affiliation{School of Physics and Astronomy, University of Edinburgh, Edinburgh EH9 3JZ, U.K.}
\author{M. A. Green}
\affiliation{School of Physical Sciences, University of Kent, Canterbury, CT2 7NH, U.K.}
\author{O. Sobolev}
\affiliation{Forschungs-Neutronenquelle Heinz Maier-Leibnitz, FRM2 Garching, 85747, Germany}
\author{Jose A. Rodriguez-Rivera}
\affiliation{NIST Center for Neutron Research, National Institute of Standards and Technology, 100 Bureau Dr., Gaithersburg, MD 20889}
\affiliation{Department of Materials Science, University of Maryland, College Park, MD  20742}
\author{H. Cao}
\affiliation{Neutron Scattering Science Division, Oak Ridge National Laboratory, Oak Ridge, TN 37831, USA}
\author{A. Daoud-Aladine}
\affiliation{ISIS Facility, Rutherford Appleton Laboratory, Didcot, U.K.}

\date{\today}
%

\begin{abstract} We construct a phase diagram of the parent compound Fe$_{1+x}$Te as a function of interstitial iron $x$ in terms of the electronic, structural, and magnetic properties.  For a concentration of $x < 10\%$, Fe$_{1+x}$Te undergoes a ``semimetal" to metal transition at approximately 70 K that is also first-order and coincident with a structural transition from a tetragonal to a monoclinic unit cell. For $x \approx 14\%$, Fe$_{1+x}$Te undergoes a second-order phase transition at approximately 58 K corresponding to a ``semimetal" to ``semimetal" transition along with a structural orthorhombic distortion.  At a critical concentration of $x \approx 11\%$, Fe$_{1+x}$Te undergoes two transitions: the higher temperature one is a second-order transition to an orthorhombic phase with incommensurate magnetic ordering and temperature-dependent propagation vector, while the lower temperature one corresponds to nucleation of a monoclinic phase with a nearly commensurate magnetic wavevector.   While both structural and magnetic transitions display similar critical behavior for $x < 10\%$ and near the critical concentration of $x \approx 11\%$, samples with large interstitial iron concentrations show a marked deviation between the critical response indicating a decoupling of the order parameters.  Analysis of temperature dependent inelastic neutron data reveals incommensurate magnetic fluctuations throughout the Fe$_{1+x}$Te phase diagram are directly connected to the``semiconductor"-like resistivity above $T_N$ and implicates scattering from spin fluctuations as the primary reason for the semiconducting or poor metallic properties.   The results suggest that doping driven Fermi surface nesting maybe the origin of the gapless and incommensurate spin response at large interstitial concentrations.\end{abstract}

\maketitle

\section{Introduction}

Understanding the nature of phase transitions in the parent superconductor compounds such as Fe$_{1+x}$Te, BaFe$_2$As$_2$, and LaOFeAs can elucidate the nature of magnetic and structural interactions in their respective superconducting phases.~\cite{Ishida_2009, Paglione_2010, Johnston_2010, Lynn09:469,Lumsden10:22,Dai12:8,Lee06:78}  The chalcogenide system Fe$_{1+x}$Te$_{1-y}Q_y$ (where $Q=$ Se or S) is particularly illuminating due to its simple crystal structure and since it allows two chemical variables to control properties: $x$ represents the amount of interstitial iron disordered throughout the crystal, and $y$ the amount of anion substitution.  Several studies have found that the two variables are correlated,\cite{Sales_2009, Hu_2009, Zajdel_2010} and new chemical methods have been successfully utilized to independently control $x$.\cite{Rodriguez_2010, Rodriguez11:2} Experimental studies on the effects of interstitial iron on magnetic, crystallographic, and transport properties of Fe$_{1+x}$Te should shed further light on the microscopic mechanism leading to superconductivity as they answer basic questions such as the universality class of the magnetic interactions between the iron cations and whether the magnetic and structural transitions are coupled.

\begin{figure}
\includegraphics[width=0.95\linewidth,angle=0.0]{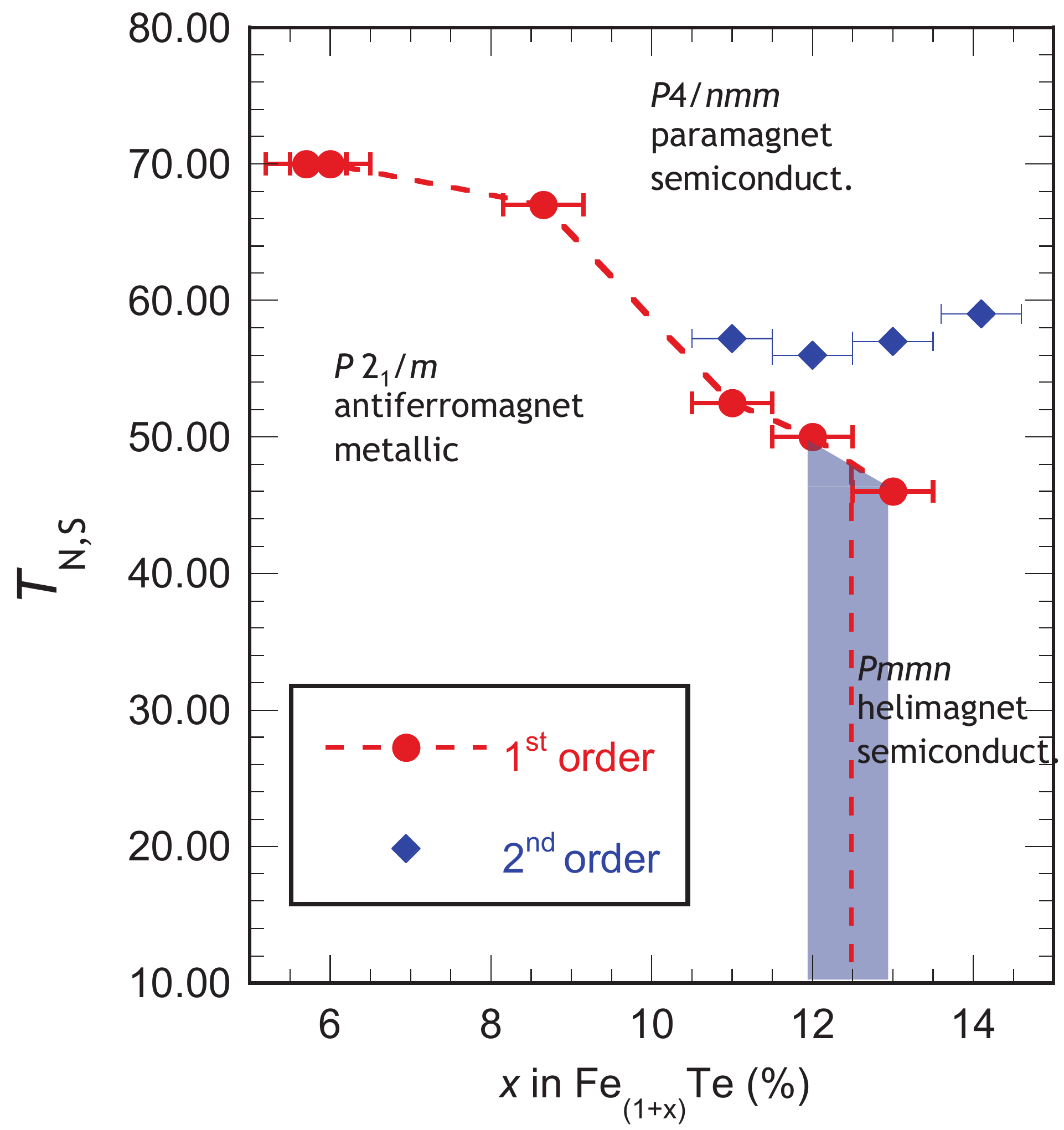}
\caption{[color online] The magnetic and structural phase transitions in Fe$_{1+x}$Te as a function of interstitial iron $x$. The first-order transition from paramagnetic to commensurate antiferromagnetic ordering corresponds to a tetragonal $P4/nmm$ to monoclinic $P2_1/m$ transition.  The second-order transition to an incommensurate helimagnet corresponds to an orthorhombic $Pmmn$ distortion.  The tricritical point is hence close to $x \approx 11-12 \%$.}
\label{FeTe_phases}
\end{figure}

In the parent compounds of iron-based superconductors, the magnetic ordering transitions are always proximate to the crystallographic phase transitions, with the notable exception being FeSe.\cite{McQueen_2009}   The nature of these transitions are useful for evaluating the dimensionality of the magnetic and structural degrees of freedom. Below 140 K, the tetragonal symmetry ($I4/mmm$) of BaFe$_2$As$_2$ is lowered to an orthorhombic one ($Fmmm$), which is concomitant with an antiferromagnetic transition.\cite{Rotter_2008, Huang_2008}  The nature (ie. first or second order) of this transition is not without controversy as some groups describe it as a continuous and second-order transition,\cite{Su_2009, Wilson_2009, Matan_2009} while others as a first-order transition.\cite{Kitagawa_2008, Kofu_2009, Sefat_2009}.  In other analogues of the 122-system such as Ca\cite{Ni_2008}, Sr,\cite{Krellner_2008, Zhao_2008b, Jesche_2008}, and Eu\cite{Jeevan_2008, Ren_2008b}, the transition has been consistently described as first-order.  Tegel \textit{et al.}, however, have argued that SrFe$_2$As$_2$ and EuFe$_2$As$_2$ are actually second-order transitions, although this conclusion will require further verification since the critical exponent for the Sr case is much smaller than that expected of a 2-D Ising system.\cite{Tegel_2008}

In contrast to the 122 family of compounds, in the REOFeAs (where RE = rare earth) system the magnetic transition $T_N$ does not coincide with the structural transition $T_S$.  Both transitions are thought to be second-order and the gap between them increases upon doping to reach the superconducting state in the phase diagram.  Interestingly, Wilson \textit{et al.} have found that although the two transitions occur at different temperatures in the same system, the order parameters corresponding to each have similar critical exponents that are close to $\beta = 0.25$.\cite{Wilson_2010}  Furthermore, in lightly doped 122-phases, the transition temperatures are also separated as in the 1111-system, and their critical exponents become more indicative of three-dimensional behavior as well.  Therefore, the question becomes whether these systems must crossover from 2D-like behavior in the parent phases to 3D behavior in order for superconductivity to be achieved.

The state of understanding these transitions in Fe$_{1+x}$Te has been complicated by the fact that the amount of interstitial iron $x$ has a significant effect on the magnetic ordering and crystallographic phase transitions.\cite{Rodriguez_2011b}  It is well established from neutron diffraction,\cite{S-Li_2009, Bao_2009} specific heat,\cite{Rossler_2011} magnetization, and transport studies\cite{Liu_2009} that for samples with $x < 10 \%$, the transition is first-order.  The lower temperature monoclinic symmetry $P2_1/m$ is not a subgroup of the higher temperature tetragonal space group $P4/nmm$.  Thus, Landau-Ginzburg theory would preclude this transition from being  second-order on grounds that the symmetry change does not occur through a single irreducible representation.  For samples with $x > 10 \%$, the nature of these transitions becomes complicated since it has been found that the magnetic propagation vector does not vary smoothly with $x$.  Instead, at a special composition of $x \approx 12 \%$, the wave vector becomes an incommensurate spin density wave, and electronic phase separation occurs.\cite{Rodriguez_2011b}  For values of $x \approx 14 \%$, the transition is better understood to be second-order from specific heat and transport studies.\cite{S-Li_2009}

In order to better understand the nature of these phase transitions in FeTe, we have performed temperature dependent neutron diffraction and inelastic neutron scattering measurements.  We then characterize the transport and specific heat of two single crystal samples to complete a picture of the phase diagram for Fe$_{1+x}$Te.  The main features describing the transition temperatures, magnetic ordering, electronic behavior, and nature/order of the phase transitions can be found in Fig. \ref{FeTe_phases}.  A tricritical-like point is found in the phase diagram where a first-order transition gives way to second-order transition as the amount of interstitial iron is increased.  In this paper, we characterize the structural, magnetic and electronic properties of this transition in detail with the goal of experimentally mapping out the properties of this transition.   Based on the coupling we observe between magnetic, structural, and electronic properties, we conclude that interstitial iron acts as a charge dopant altering the Fermi surface nesting and dramatically changing the metallic properties. 

The paper is divided into three sections with the first outlining the experiments, the second discussing the results throughout the phase diagram, and the third providing a summary and conclusions.  Neutron elastic and inelastic scattering results outline the changes observed in the magnetic structure and excitations in the three regions of the phase diagram; 1) commensurate antiferromagnetic region ($x<$11\%); 2) near the tricritcal-like point  ($x\sim$11\%); 3) heavily iron doped samples  ($x>$11\%).  Given that the tricritical point separates a phase described by a commensurate wavevector from an incommensurate one, it is defined as a Lifshitz point.  Indeed, other transition metal systems have expressed such a point in their rich magnetic phase diagrams.  For example, the magnetic field vs. temperature phase diagram in MnP reveals such a Lifshitz point at the intersection of para, ferro, and complex incommensurate magnetism.~\cite{Becerra:1980wc,Shapira:1981uf,Yoshizawa:1985vj,Zieba:1992vn,Zieba:2000te}

\section{Experimental}

The powder samples used for these studies were prepared by mixing stoichiometric ratios of iron and tellurium powders and heating them in evacuated quartz glass ampoules up to 400 $^{\circ}$ C for 12 hours followed by 700 $^{\circ}$ C for several days before furnace cooling.  For samples with $x > 12 \%$, a second reaction with more iron powder was performed, presumably due to some iron loss from reaction with the quartz glass walls.  Single crystals were grown by the Bridgemann method, starting from a melt temperature of 825 $^{\circ}$ C and cooling rate of 3 $^{\circ}$ C/hour.   Specific heat measurements were performed on small slices taken from the large neutron single crystals using a laboratory based PPMS system from Quantum design.  Resistivity was obtained using the 4-probe lock-in technique.  Further details on the sample preparation and determination of interstitial iron can be found in Ref \onlinecite{Rodriguez_2011b}.    

For the single crystal diffraction, we utilized the HB-3A four-circle diffractometer at the High-flux Isotope Reactor (HFIR) at Oak Ridge National Laboratory (Oak Ridge, USA).   Both nuclear and magnetic reflections were measured using the Si(220) monochromator with a wavelength of 1.5424 \AA.   In order to follow the subtle phase separation in low-interstitial iron samples, the powder diffraction measurements were performed at the high-resolution powder diffractometer (HRPD) at the ISIS spallation source facility at Rutherford Appleton Laboratory (Didcot, UK).   The high resolution capabilities provided by HRPD combined with the direct coupling between neutrons and the magnetic spin allowed both magnetic and structural properties to be tracked simultaneously with temperature.  This was important for mapping out the critical properties of both structural and magnetic order parameters simultaneously in a single experiment.

Neutron inelastic experiments investigating the critical scattering near the magnetic ordering temperature were carried out using the MACS cold triple-axis spectrometer (NIST,USA).    Instrument and design concepts can be found elsewhere.~\cite{Rodriguez_2008}  Constant energy planes were scanned by fixing the final energy at E$_{f}$=3.6 meV using the 20 double-bounce PG(002) analyzing crystals and detectors and varying the incident energy defined by a double-focused PG(002) monochromator.  Each detector channel is collimated using 90$'$ Soller slits before the analyzing crystal.  Full maps of the spin excitations in the (H0L) scattering plane, as a function of energy transfer, were then constructed by measuring a series of constant energy planes.  All of the data have been corrected for $\lambda/2$ contamination of the incident-beam monitor and an empty cryostat background has been subtracted.

To probe the higher energy magnetic fluctuations, and in particular the temperature dependence of the high temperature incommensurate fluctuations, we performed experiments at the PUMA thermal triple-axis spectrometer (FRM2, Germany).~\cite{Link00:276}   A vertically focusing and horizontally flat PG(002) monochromator were used with a horizontally flat PG(002) analyzer.  Soller slit collimators were used and the sequence were fixed at $40'-mono-40'-S-open-analyzer-open$.  The experiments used a fixed final energy of 13.5 meV and a PG filter was placed on the scattered side to remove higher order contamination from the monochromator.  The counting times were corrected for higher order contamination of the incident beam monitor as described elsewhere.~\cite{Shirane:book,Stock04:69}

Control of temperature is a key factor in measuring critical properties in diffraction experiments.  We present critical scattering data from HRPD (ISIS, UK) and HB-3A (HFIR,Oak Ridge).  On HRPD, powder samples were sealed in a thin plate geometry with thermometers mounted on the top and bottom of the plates.  Temperature dependent patterns were recording for approximately 1.5 hours each, in 2 K steps, with a 5 minute wait for changing temperature.   On HB-3A, the sample was heated from 4 to 46 K in 1 K steps followed by 0.5 K steps up to 68 K and then 1 K steps were used up to 80 K.  A 5 minute wait between temperatures was used.  For both HB-3A and HRPD, a temperature stability of better than 0.15 K was achieved for the measurements.  In the analysis presented, the range over which the critical exponents were fitted are shown in the figures and in the range of $t \equiv {{|T-T_{c}|} \over {T_{c}}} < 0.2$.    

There has been considerable work reported in the analysis and behavior of critical properties near phase transitions in a variety of systems.~\cite{Collins:book,Stanley:book,Stanley99:71}  The response and the critical properties of the iron based superconductors are exceptionally complex owing to the competition between magnetic, structural, and electronic (including superconducting) degrees of freedom. The critical magnetic and structural scattering (from x-ray and neutron experiments) from BaFe$_{2}$As$_{2}$ and LaFAsO have been described in the literature in terms of a single critical exponent (Ref. \onlinecite{Wilson_2010}) and for simplicity, and to facilitate a direct comparison with these previous works, we have followed this analysis and only considered the critical scattering in terms of a single exponent.  Over the the range of $t$ probed and analyzed, Ref. \onlinecite{Pajerowski13:87}, show that the critical dynamics can be described within a single exponent.

\section{Results}

In this section we outline the experimental results in three key regions of the phase diagram illustrated in Fig. \ref{FeTe_phases}.  The first section describes the critical properties for concentrations below $x<$ 11\% which is characterized by commensurate collinear antiferromagnetic order.  The second section discusses the other extreme of the phase diagram for concentrations $x>$ 11\%.  This region of the phase diagram is described by spiral magnetic order.  The third section discusses the properties near the critical concentration of $x\sim$ 11\%, which is where a tricritical point exists in the phase diagram.

\subsection{First-order phase transition and magnetic incommensurate fluctuations for $x < 10 \%$}

\begin{figure}[t]
\includegraphics[width=0.95\linewidth,angle=0.0]{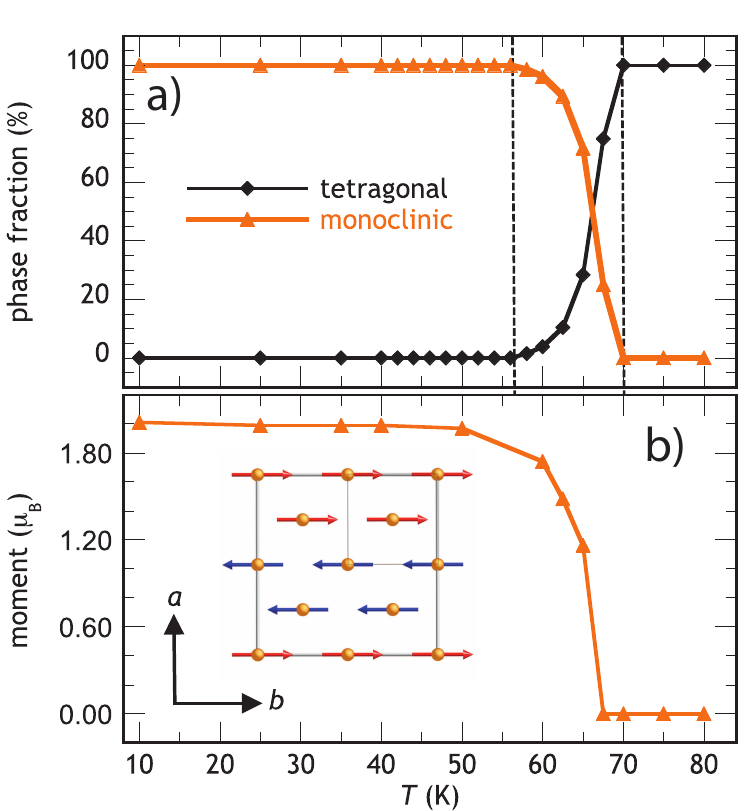}
\caption{[color online]  (a) Temperature evolution of the phase fraction in Fe$_{1.057(7)}$Te from neutron powder diffraction data.  The mixed tetragonal and monoclinic phases for the $T$-range enclosed by the dashed lines is clear evidence of the first-order nature of this transition. (b) The size of the magnetic moment per iron cation as a function of temperature.  Comparison of the moment size with the monoclinic phase fraction demonstrates how $T_N$ and $T_S$ are simultaneous and sudden in this composition. Inset shows the antiferromagnetic unit cell within the $ab$-plane and the moment direction.  In this figure and all others in this paper, when not indicated otherwise, errorbars are the size of the data points and represent $\pm 1 \sigma$.}
\label{commens_phase}
\end{figure}

As previously found, compositions for low-interstitial iron lead to a first-order magnetic and structural transition at approximately 70 K.   As shown in Fig. \ref{FeTe_phases}, the purely first-order nature of these transitions can be found for samples with as much interstitial iron as $x \approx 9 \%$.  This transition was explored in detail using a powder sample of Fe$_{1.057(7)}$Te and high resolution neutron diffraction facilitated by HRPD (ISIS, UK). The results revealed a change in crystal symmetry from monoclinic $P2_1/m$ to tetragonal $P4/nmm$ and the use of a high resolution neutron diffractometer allowed the magnetic ordering and the structural transitions to be monitored simultaneously.  Based on a comparison of panels $a)$ and $b)$ in Fig. \ref{commens_phase}, the magnetic transition and transition to a monoclinic unit cell occur at nearly the same temperature indicating a strong magnetoelastic coupling in this material.

\begin{figure}[t]
\includegraphics[width=0.9\linewidth,angle=0.0]{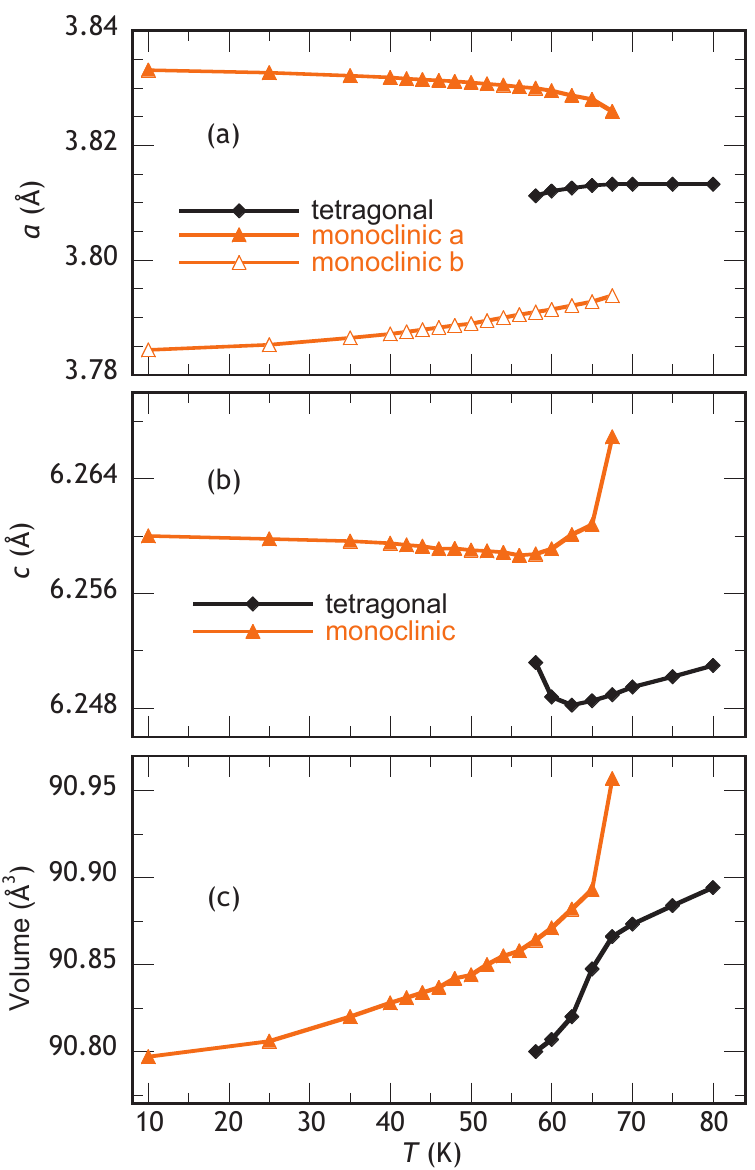}
\caption{[color online]  Temperature evolution of the lattice parameters for Fe$_{1.057(7)}$Te from neutron powder diffraction data.  In (a) the a- and b-lattice parameters corresponding to the in plane Fe-Fe distance; in (b) the c-parameter, and in (c) the discontinuous volume change indicative of a first-order phase transition.}
\label{commens_lattice}
\end{figure}

The phase transition is further explored in Fig.  \ref{commens_lattice}, where the lattice constants and volume are plotted.  Several features of these results confirm the structural transition $T_S$ to be first-order:  first, the large region of phase coexistence of the monoclinic and tetragonal phases (Fig. \ref{commens_phase}), and second, the discontinuous change in the lattice parameters and lattice volume (Fig. \ref{commens_lattice}$a-c$).   The coupling between commensurate collinear magnetic order and the monoclinic low temperature phase has been theoretically analyzed in terms of Landau theory in Ref. \onlinecite{Paul11:83}, where strong magnetoelastic coupling was noted.  A strong coupling has also been theoretically suggested through a correlation between the structural and orbital properties.~\cite{Turner_2009,Lee_2010} Photoemission studies have not been able to observe a nesting wave vector for interstitial iron concentrations in this range implying that the collinear magnetic order is not associated with a Fermi surface instability.~\cite{Xia09:103}

\begin{figure}[t]
\includegraphics[width=0.95\linewidth,angle=0.0]{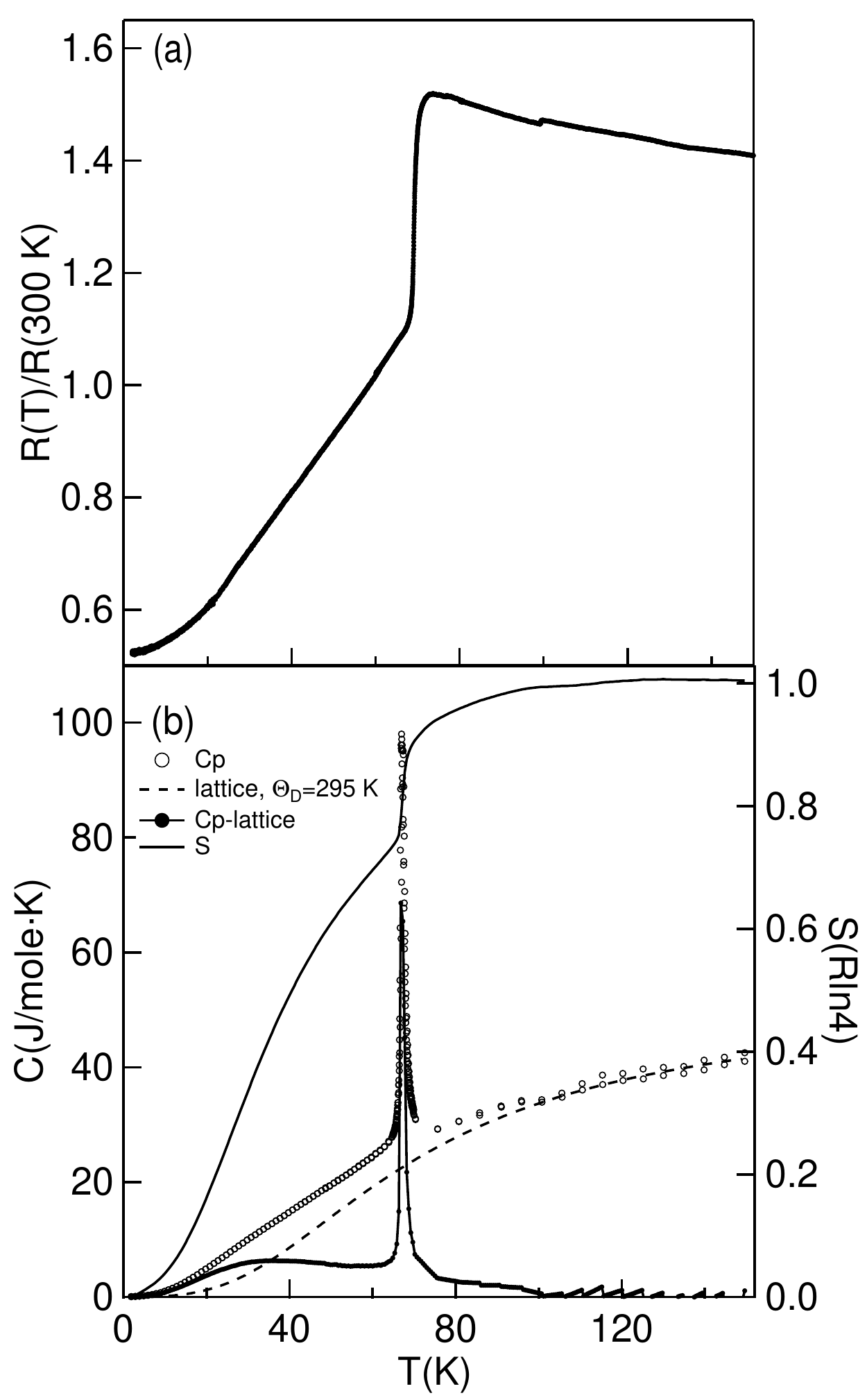}
\caption{$a)$ The in-plane resistivity and $b)$ heat capacity as a function of temperature for Fe$_{1.057(7)}$Te.  The drop in resistivity marks a metal-to-semiconductor transition upon warming, which corresponds to a large increase in the specific heat. Note the slightly jump at $\sim$ 100 K is an artifact in the data collection.  The solid filled circles display the estimated magnetic contribution to the specific heat - namely $C_{p}-C_{lattice}$.  The sawtooth structure at high temperature in the estimated magnetic contribution to the specific heat is a result of using nearest neighbor interpolation routine.  The entropy is plotted in $b)$ in units of $R \ln 4$ and shows that the fully entropy for a $S=3/2$ moment is acquired above T$_{N}$.}
\label{commens_transport}
\end{figure}

Specific heat and electrical resistivity measurements on a single crystal with $x \approx 6\%$ (taken from the same batch studied with neutrons in Ref. \onlinecite{Stock11:84} and on HRPD shown above) corroborate the first-order nature of $T_N$ and $T_S$ (Fig. \ref{commens_transport} with both data sets taken on warming).  Interestingly, the large peak in the specific heat seems to precede the sudden change in resistivity by $\approx 2$ K and the resistivity shows a sudden drop at $\sim$ 70 K, also reflecting the first order nature of the transition.  Overall, the transition appears to be a semiconductor-to-metal transition based upon resistivity (Fig. \ref{commens_transport} $a)$).  The term semiconductor or semimetal is used here as the resistivity at high temperatures does not vary strongly and does not indicate a low temperature divergence as might be expected for an insulator. 

The magnetic entropy was extracted from the specific heat and is shown in Fig. \ref{commens_transport} $b)$.  A lattice contribution was extracted by fitting a Debye model ($C_{p}=3\times 9 R\left( {T\over\Theta_{D}}\right)^{3}\int_{0}^{\Theta_{D}/T} dx {{x^{4}e^{x}}\over{(e^{x}-1)}}$) with a fitted Debye temperature of $\Theta=295 \pm 2 K$, and subtracted from the total specific heat to obtain the purely magnetic contribution.  The magnetic entropy sharply approaches $Rln4$, expected for a $S={3\over2}$ system in agreement with previous analysis.~\cite{Zaliznyak11:107,Zaliznyak12:85}  The abrupt recovery of all of the magnetic entropy at T$_{N}$ indicates electronically localized behavior for small values of interstitial iron below the critical concentration of $x\sim 11\%$.  

Based on a comparison with the resistivity in Fig. \ref{commens_transport} and the high resolution diffraction data obtained on HRPD in Fig. \ref{commens_phase} and \ref{commens_lattice}, the initial drop in resistivity at 70 K can be associated with onset of commensurate magnetic order and the change in unit cell shape.  The correlation between the magnetic order and spin fluctuations to the change in resistivity is made later on in the paper.  

\begin{figure}[t]
\includegraphics[width=8.7cm] {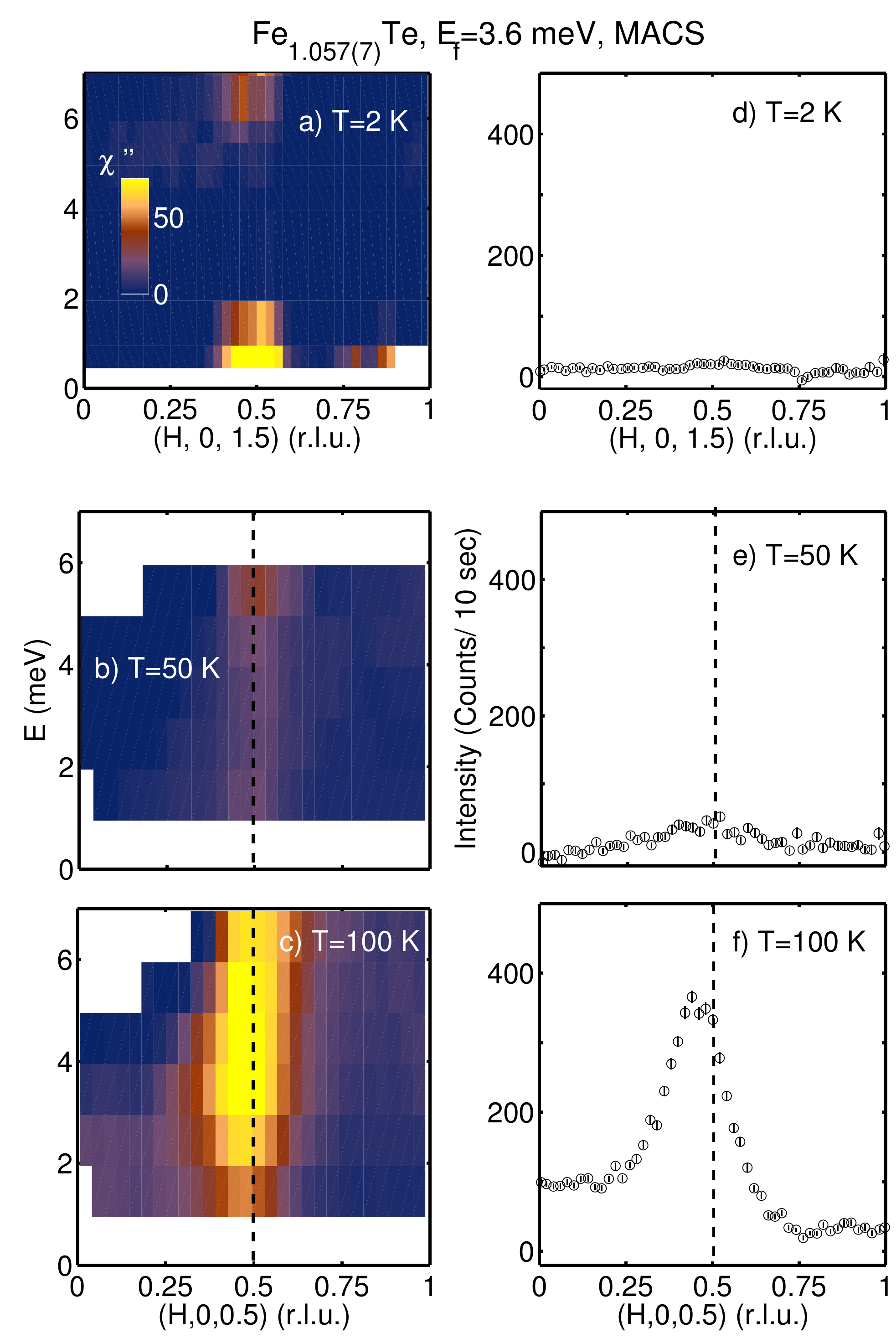}
\caption{\label{constantQ_figure} [color online] Constant-Q scans through the magnetic correlations in Fe$_{1.057(7)}$Te. $a-c)$ show constant Q slices taken over (H,0,L$\pm$ 0.15). $d-f)$ illustrate constant E=2.0 meV cuts integrating around L $\pm$ 0.15. Note that the color images show $\chi$$''$ as discussed in the text.}
\end{figure}

Having shown that the commensurate magnetic structure is determined by a combined first order magnetic and structural transition, we now study the fluctuating magnetic critical scattering and how it changes in energy and momentum on passing through this transition temperature.  Fig. \ref{constantQ_figure}  summarizes the magnetic response near the first order transition and plots constant-Q slices integrating over $\vec{Q}$=({${1\over 2}\pm 0.05 $,0,${1\over 2}\pm 0.15$}).  Fig. \ref{constantQ_figure} shows a plot of the intensity and imaginary part of the susceptibility $\chi''$ which are related via $I(\vec{Q},E) \propto S(\vec{Q},E)\equiv{1 \over \pi} [n(E)+1] \chi'' (\vec{Q},E)$.   At low temperatures, the magnetic intensity and susceptibility is gapped as reported previously.~\cite{Stock11:84}  At 50 K, weak momentum broadened magnetic fluctuations appear at intermediate energy transfers.  At high temperatures of 100 K, above the magnetic ordering transition temperature, the energy spectrum is replaced by broad over damped fluctuations which are located at a slightly incommensurate position (panel $c)$).  The temperature range correspond to where weak incommensurate order was reported in some concentrations of Fe$_{1+x}$Te, with $x<0.12$.~\cite{Rodriguez_2011b}  Inelastic incommensurate fluctuations with a similar wavevector were also reported for similar interstitial iron concentrations.~\cite{Parshall12:85}   Constant energy cuts along $\vec{Q}$=$(H,0,{1 \over 2} \pm 0.15)$ are presented in panels $d)-f)$ and show the increase of incommensurate magnetic fluctuations above T$_{N}$.  The scattering near H=0 are due to phonons and increase with both temperature and momentum transfer.

\begin{figure}[t]
\includegraphics[width=8.8cm] {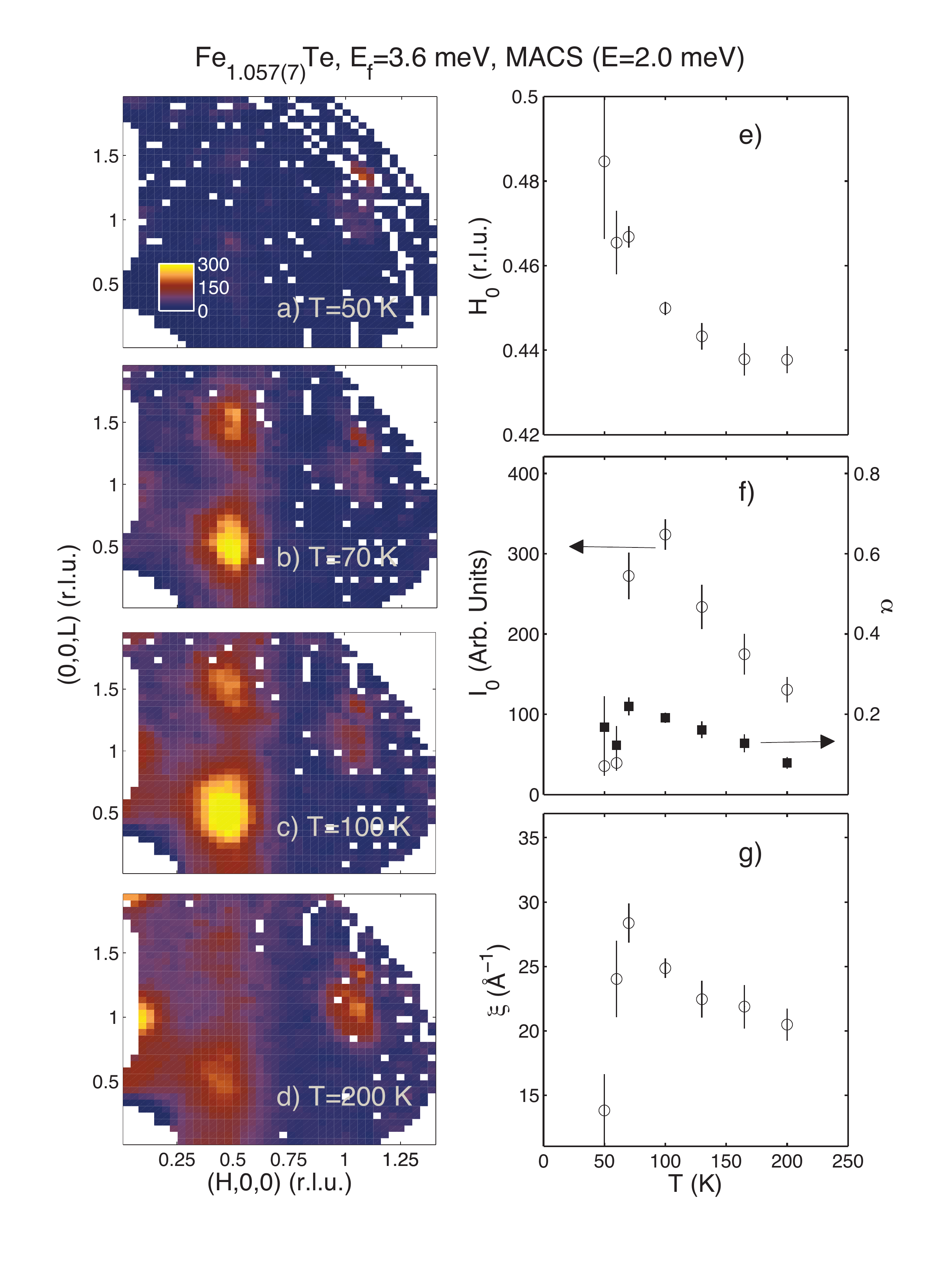}
\caption{\label{constantE_figure} [color online] Constant E=2.0 meV scans through the magnetic correlations in Fe$_{1.057(7)}$Te. $a-d)$ Show constant energy slices at a series of temperatures.  $e)$ displays the fitted position along the (H,0,0.5) direction based upon cuts through similar scans displayed in $a-d)$.  $f)$ shows the peak intensity ($I_{0}$) as a function of temperature and the parameter $\alpha$ (described in the text) which parameterizes the correlations along $L$.  $g)$ plots the dynamic correlation length taken at E=2.0 meV.  The intensity around $\vec{Q}$=(1,0,1) is due to a phonon.}
\end{figure}

Figure \ref{constantE_figure} plots the momentum dependence of the scattering in the (H0L) scattering plane at an energy transfer of 2.0 meV.   Panels $a-d)$ illustrate constant energy slices at E=2.0 meV for a series of temperatures.  The results prove that the fluctuations are gapped in the magnetically ordered state, and that the strong low energy and incommensurate fluctuations are present near and above T$_{N}$.  Panels $e-f)$ show a Lorentzian squared fit to the data taken at each temperature.  A Lorentzian squared was chosen as the integral is finite in two dimensions and also can be related to the presence of random fields.  The position along (H,0,${1\over 2}$) are plotted in panel $f)$ where a trend towards to the commensurate H=0.5 point at T$_{N}$ is demonstrated.  Fig. \ref{constantE_figure} $f)$ illustrates the intensity ($I_{0}$) from scans along $H$ as a function of temperature and show a gradual increasing trend in intensity at T$_{N}$.  However, the results are far from critical in nature.  Panel $f)$ also displays the results of fitting the $L$ dependence to $A(1+2\alpha \cos(2\pi L+\pi))$ also exhibiting an increase of correlations between the FeTe layers, as measured by the parameter $\alpha$, as T$_{N}$ is approached.  Panel $g)$ displays a plot of the dynamic correlation length, extracted from H-scans at E=2.0 meV, as a function of temperature showing a gradual increase, and then a sharp drop in the correlation length at T$_{N}$.  While we emphasize that this is not a true equal-time correlation length that can be used to characterize magnetic transitions, the results again show that the correlations are far from critical and the lengthscales are small $\sim$ 20-30 \AA.  

\begin{figure}[t]
\includegraphics[width=8.5cm] {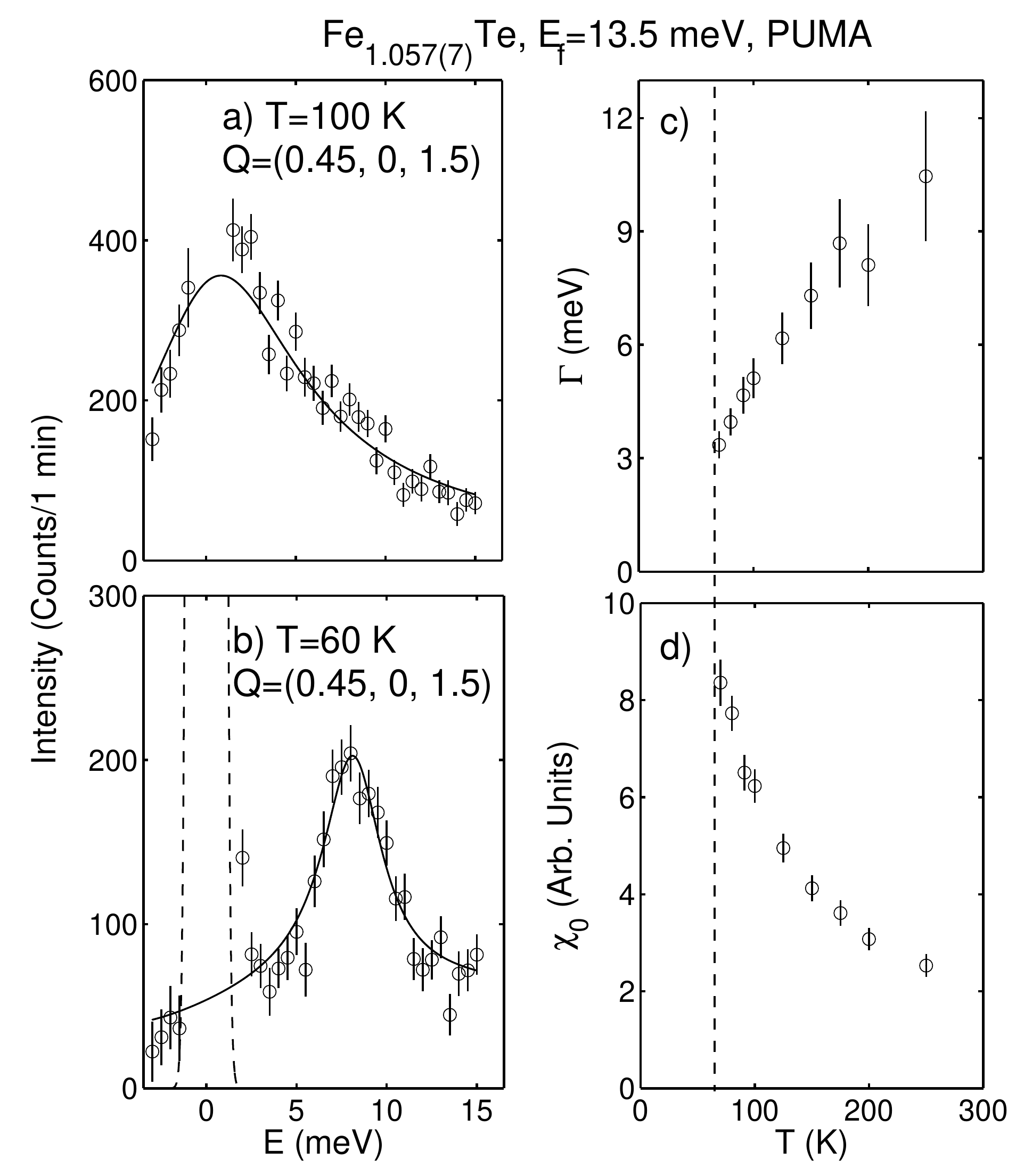}
\caption{\label{puma_figure} Constant-Q scans taken on the PUMA thermal triple-axis spectrometer.  $a-b)$ plot energy scans taken at $\vec{Q}$=(0.45,0,0.5) at 100 and 60 K.  $c)$ illustrates the half-width ($\Gamma$) as a function of temperature and $d)$ shows the temperature dependent parameter $\chi_{0}$. }
\end{figure}

We now focus on the temperature dependence of the magnetic fluctuations in weakly interstitial iron doped Fe$_{1.057(7)}$Te above T$_{N}$ in the paramagnetic phase where the structure is tetragonal.  To investigate the temperature dependence of the incommensurate fluctuations we used the PUMA (FRM2, Germany) thermal triple-axis which can access higher energy transfers than the cold neutron spectrometer MACS.  Examples of constant-Q scans are displayed in Fig. \ref{puma_figure} $a-b)$, above and below T$_{N}$, and are consistent with the slices presented at lower energies from MACS.  The data have been corrected for a background taken at $\vec{Q}$=(0.75,0,1.5) where the magnetic scattering is strongly suppressed (Fig. \ref{constantE_figure} $a-d)$).   Panel $a)$ plots a scan at T=100 K and is representative of the typical structure of the magnetic fluctuations above T$_{N}$.  The magnetic scattering is over damped and gapless, in constrast to the gapped structure found in the ordered state (Fig. \ref{constantQ_figure} $a)$).  At temperatures of 60 K, just below T$_{N}$, the magnetic spectrum is gapped though with some low-energy fluctuations present as demonstrated in Fig. \ref{constantE_figure}.  The stark difference between the two lineshapes is indicative of the first-order transition.  

The energy dependence at each temperature above T$_{N}$ was fit to the following relaxational form describing over damped spin excitations on a single relaxational energy scale,

\begin{eqnarray}
I(E)\propto \chi_{0} [n(E)+1] {E\over {1+\left(E/\Gamma\right)^2}}.
\label{equation_chi} 
\end{eqnarray}

\noindent where $\chi_{0}$ is proportional to the real part of the susceptibility, $\Gamma$ is related to the lifetime of the spin excitations $1/\tau$, and $[n(E)+1]$ is the thermal population or Bose factor.  The $\chi_{0}$ and $\Gamma$ parameters are plotted in Fig. \ref{puma_figure} $c)$ and $d)$.  The results show a decrease of $\Gamma$, indicative of a slowing of the spin fluctuations as the transition temperature is approached.  Similarly, $d)$ shows an increase of the antiferromagnetic susceptibility presenting a gradual and increasing trend with decreasing temperature.  Neither the lifetime ($\tau)$ nor the susceptibility ($\chi_{0}$) fully diverge at T$_{N}$ and their fundamental change across the transition temperature is indicative of discontinuous change in properties.  

The neutron inelastic scattering results indicate a competition between gapped commensurate and gapless incommensurate fluctuations above T$_{N}$.  Above T$_{N}$, short-range incommensurate fluctuations at low-energies exist, which gradually approach the commensurate point.  Below T$_{N}$, these are suppressed at the first order transition and replaced by gapped magnetic fluctuations.   It is interesting to compare these results with the electrical resistivity measured on the same crystals (Fig. \ref{commens_transport}).  At T$_{N}$, a transition from a semiconducting to a metallic states exists.  At the same time, incommensurate fluctuations are suppressed, therefore indicating a direct connection between the low-energy incommensurate fluctuations and the semiconducting resistivity.   We address this point later on when we isolate the purely spin component of the resistivity and compare results from different points in the Fe$_{1+x}$Te phase diagram. 

\subsection{Second-order phase transitions and 2D critical fluctuations for $x \approx 14\%$}

Having discussed the critical properties of lightly Fe doped Fe$_{1+x}$Te, we now focus on the other part of the phase diagram for large amounts of interstitial iron.  When the maximum amount of interstitial iron is filled ($x \approx 14 - 16 \%$) in Fe$_{1+x}$Te, the nature of both the magnetic and structural transitions become second-order and the magnetic fluctuations become gapless in energy.  The electronic properties reflect a semi-metallic state for all temperatures above and below the magnetic and structural transitions.

We first discuss the structural properties for Fe$_{1.141(5)}$Te which is located well right of the critical $x\sim$ 11\% concentration illustrated in Fig. \ref{FeTe_phases} and in the magnetic incommensurate region of the phase diagram. As discussed previously in our work on the structural and static magnetic properties, upon cooling, the tetragonal phase, $P4/nmm$, distorts to an orthorhombic phase, $Pmmn$.  Note that $Pmmn$ is a maximal subgroup of $P4/nmm$ therefore allowing the transition to be second order by Landau theory.  

\begin{figure}[t]
\includegraphics[width=0.93\linewidth,angle=0.0]{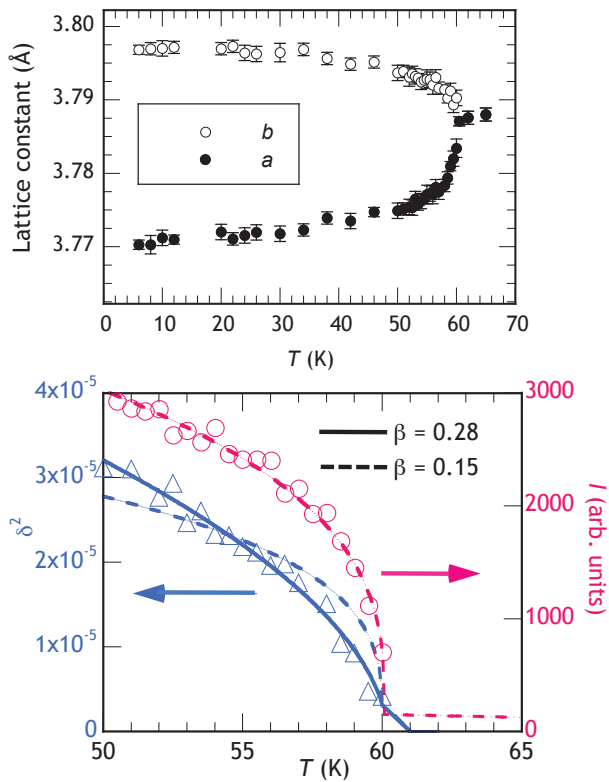}
\caption{[color online]   (a) Temperature evolution of the orthorhombic to tetragonal distortion in Fe$_{1.141(5)}$Te as measured by the (040) reflection form single crystal neutron diffraction experiments.  (b) Comparison of the structural order parameter, $\delta ^2$, and the magnetic order parameter as determined by measuring the (0.38 0 0.5) magnetic reflection. The critical exponent was fit only over the temperature range shown by allowing the transition temperature and exponent to vary.}
\label{incommens_order}
\end{figure}

This distortion was followed as a function of temperature by measuring the (040) reflection of a single crystal of Fe$_{1.141(5)}$Te using the single crystal neutron diffractometer HB-3A at HIFR (Oak Ridge).  The results are illustrated in Fig. \ref{incommens_order}.  The lattice constants are plotted in panel $a)$ and the structural and magnetic order parameters are illustrated in panel $b)$.

The structural order parameter is defined here as $\delta/2 = (a-b)/(a+b)$, as done previously for following the orthorhombic distortion in FeAs-based systems.  From a fit to the form $\delta^{2}\propto (T-T_{c})^{2\beta}$ we derive an exponent of $\beta=0.28(5)$ for the structural order parameter.  The large errorbar was determined by fitting the data set over different temperature ranges.  To allow a direct comparison, we have chosen to present the data fit over the same temperature range presented for the magnetic intensity discussed below.  3D Ising predicts $\beta$=0.326 while 3D Heisenberg and X-Y predict 0.367 and 0.345 respectively.  While the data does not implicate a single universality class, it is clear that the experimental exponent for large interstitial iron concentrations reflect more 3D-like behavior than 2D with the dashed line in Fig. \ref{incommens_order} showing a fit forced to the data with the 2D $\beta$=0.15.  The order parameter squared, $\delta^2$, can then be compared to the intensity of the magnetic peak centered at approximately $\vec{Q}$=(0.38, 0, 0.5).  Since the intensity of the magnetic Bragg peak from neutrons is proportional to the magnetization squared, and hence is the magnetic order parameter, we can compare the two order parameters directly  as done in Fig. \ref{incommens_order} b).  Even though $T_N$ and $T_S$ are concomitant and both represent continuous transitions, it is apparent that the temperature dependence of $\delta$ and $M$ are different and this point is now discussed further.  

\begin{figure}[t]
\includegraphics[width=8.5cm] {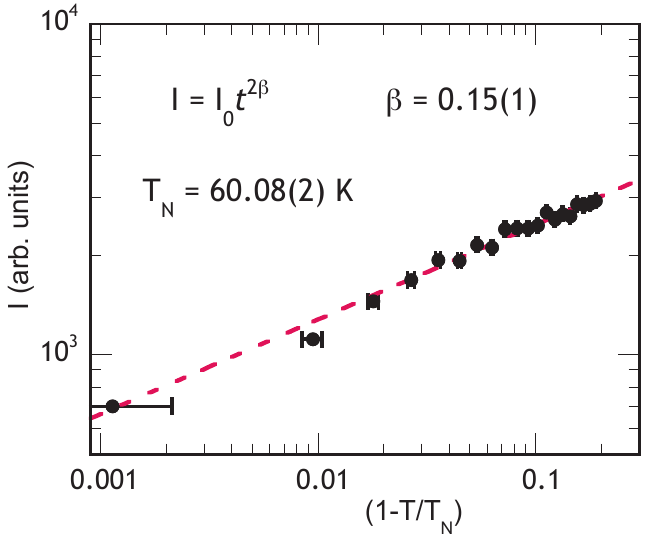}
\caption{ [color online]  A power law fit to the intensity of the magnetic Bragg reflection in Fe$_{1.141(5)}$Te yielding an exponent of $\beta$=0.15(1), consistent with 2D-Ising behavior.}
\label{incommens_exponent}
\end{figure}

A power law fit to $M^2$ versus reduced temperature $t$ ($=1-T$/$T_N$) afforded a critical exponent $\beta$ of 0.15(1) (Fig. \ref{incommens_exponent}).   The value for $\beta$ is within error to that of the 2D-Heisenberg universality class found in for the magnetic order parameter in BaFe$_{2}$As$_{2}$ and SrFe$_{2}$As$_{2}$ compounds.  The value also agrees with exponents derived in classic 2D transition metal magnets like in K$_{2}$NiF$_{4}$~\cite{Birgeneau_1983} as well as the parent FeAs compound.~\cite{Rodriguez_2011b}  The 2D critical fluctuations are also consistent with earlier reported inelastic studies showing weak correlations between the FeTe planes in Fe$_{1.141(5)}$Te, an indication of 2D fluctuations.~\cite{Stock11:84}   The value for $\beta$ contrasts with that found in doped pnictide compounds, where the critical fluctuations cross over to to more 3D type behavior upon approaching the transition to a high temperature superconductor.  Indeed, doping charge carriers through interstitial iron in Fe$_{1+x}$Te appears to retain the 2D character for the magnetic critical properties.

Another difference therefore between Fe$_{1+x}$Te and the pnictides is the divergent critical properties between the structural and magnetic order parameters.  As pointed out by Wilson \textit{et al.} ~\cite{Wilson_2010} single layer pnictides display similar critical properties for structural and magnetic order parameters as evidenced by LaFeAsO~\cite{Yan09:95,Maeter90:80}.  Heavily doped Fe$_{1+x}$Te, however, shows 3D critical properties for the structural fluctuations, while 2D for the magnetic.  This suggests a possible decoupling of the two order parameters with increased doping through interstitial iron.  This also contrasts with the parent phase BaFe$_{2}$As$_{2}$ compounds where uniaxial strain demonstrates a coupling between structural and magnetic order parameters.~\cite{Dhital12:108}  In SrFe$_{2}$As$_{2}$, the structural and magnetic transitions are coincident and the order parameters are tracked with a single exponent indicative of strong coupling.~\cite{Zhao_2008b, Jesche_2008}  Recent work on doped BaFe$_{2}$As$_{2}$ has found that while the two temperature scales for magnetic and structural transitions separate with charge doping, they converge again near optimal dopings for superconductivity.~\cite{Lu13:110}

\begin{figure}[t]
\includegraphics[width=0.95\linewidth,angle=0.0]{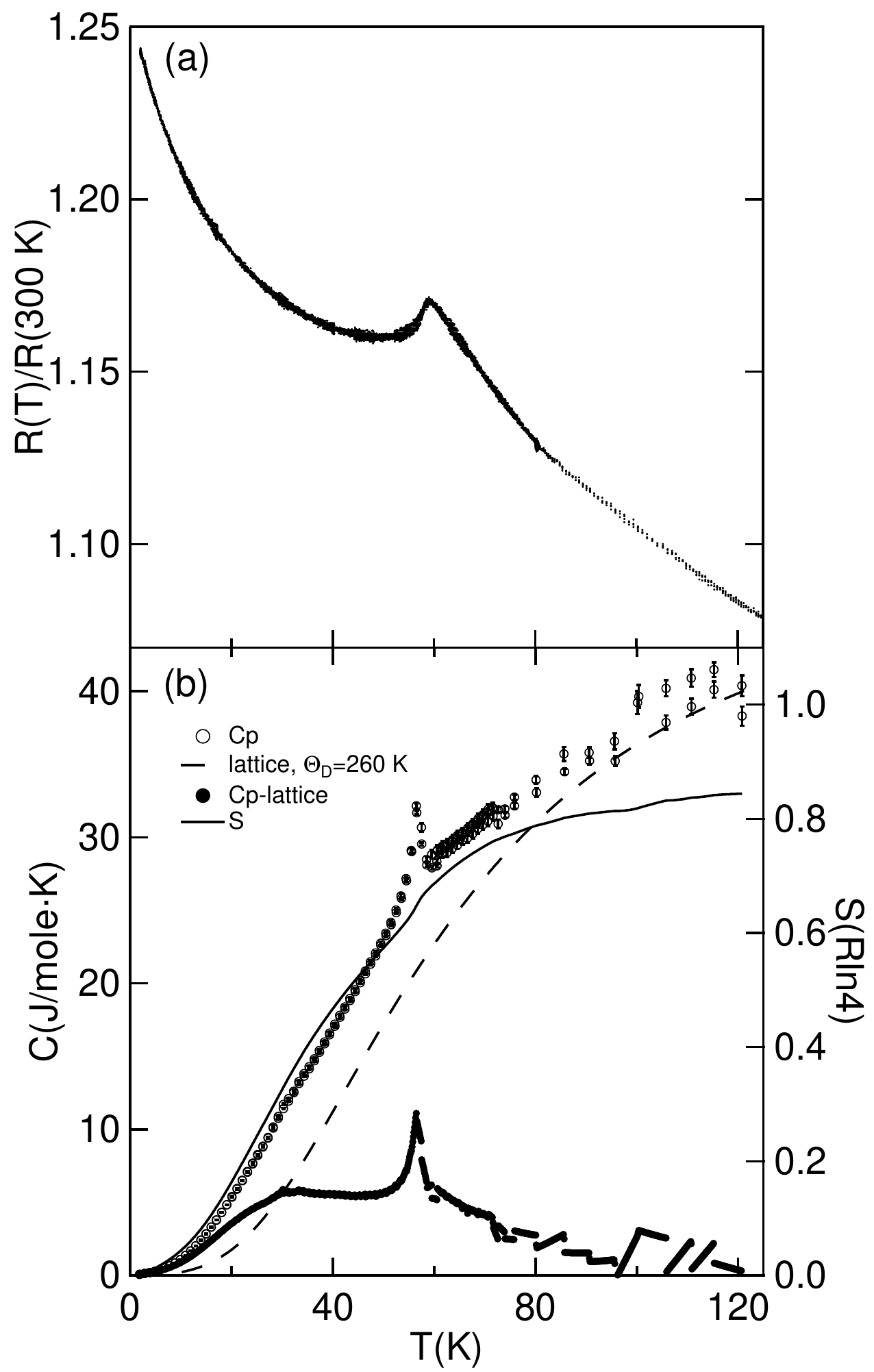}
\caption{ Temperature dependent in-plane electrical resistivity $a)$ and specific heat $b)$ for Fe$_{1.141(5)}$Te.  The entropy in units of $Rln4$ and, in contrast to the commensurate sample, shows only a gradual increase above T$_{N}$ and never reaches the full saturation value for $S=3/2$.  The estimated magnetic contribution to the heat capacity is shown by the filled circles ($C_{p}-C_{lattice}$).  The sawtooth structure to the magnetic portion at high temperatures is a result of using nearest neighbor interpolation from the data.}
\label{incommens_transport}
\end{figure}

Having elucidated the nature of the magnetic and structural second order transition, we now discuss the transport properties in this region of the phase diagram.  In Fig. \ref{incommens_transport}, the heat capacity is displayed in panel $b)$ and shows a peak at the structural and magnetic transitions at $\sim$ 60 K.  In a similar manner to the commensurate data described above (Fig. \ref{commens_transport}), the lattice contribution to the heat capacity was obtained by fitting a Debye model with $\Theta=260 \pm 2 K$.  This was then used to extract the purely magnetic contribution to the specific heat.  A broad hump in the data at $\sim$ 30 K is consistent with the broad peak at $\sim$ 3 meV observed in the magnetic fluctuations at low temperatures.~\cite{Stock11:84}   In contrast to the commensurate samples with small interstitial iron concentrations, the magnetic entropy of $R \ln 4$ is not fully recovered even at 120 K - nearly twice T$_{N}$.  This is in contrast to the sharp recovery of the full magnetic entropy displayed in Fig. \ref{commens_transport} and is reminiscent of itinerant and heavy fermion systems such as Ce(Rh,Ir)In$_{5}$ and Ce$_{3}$Pt$_{4}$In$_{13}$ where the localized moments are screened by conduction electrons.~\cite{Hundley2001:65,Pagliuso2001:64}  The results are therefore suggestive that with increasing charge doping through interstitial iron, Fe$_{1+x}$Te crosses over from localized moment behavior to electronically itinerant.  The opposite trend is observed in Cr-doped BaFe$_{2}$As$_{2}$ where susceptibility measurements suggest a crossing over to a more localized magnet with increased charge doping.~\cite{Clancy12:85} 

The resistivity is in complete contrast to the commensurate data (Fig. \ref{commens_transport}).  While showing a transition at $\sim$ 60 K, the resistivity remains semi metallic down to the lowest temperature.  The data is not divergent at low temperatures and therefore not insulating.  We discuss a mechanism for this semiconducting, or sometimes referred to as a poorly metallic, properties in relation to the spin fluctuations later in the text.

The magnetic and structural order parameters, along with the transport and specific heat data, will now be placed in context to the inelastic neutron data.  The temperature dependence of the magnetic excitations reflect the critical scattering for the transition to incommensurate magnetic order.  The results above showed that for the weakly iron doped side of the phase diagram, the excitation spectrum consists of gapped commensurate fluctuations.  On approaching T$_{N}$, the gapless incommensurate fluctuations dominate the energy spectrum.  We now compare the results to the iron rich region of the phase diagram by investigating the critical fluctuations in Fe$_{1.141(5)}$Te.

\begin{figure}[t]
\includegraphics[width=8.6cm] {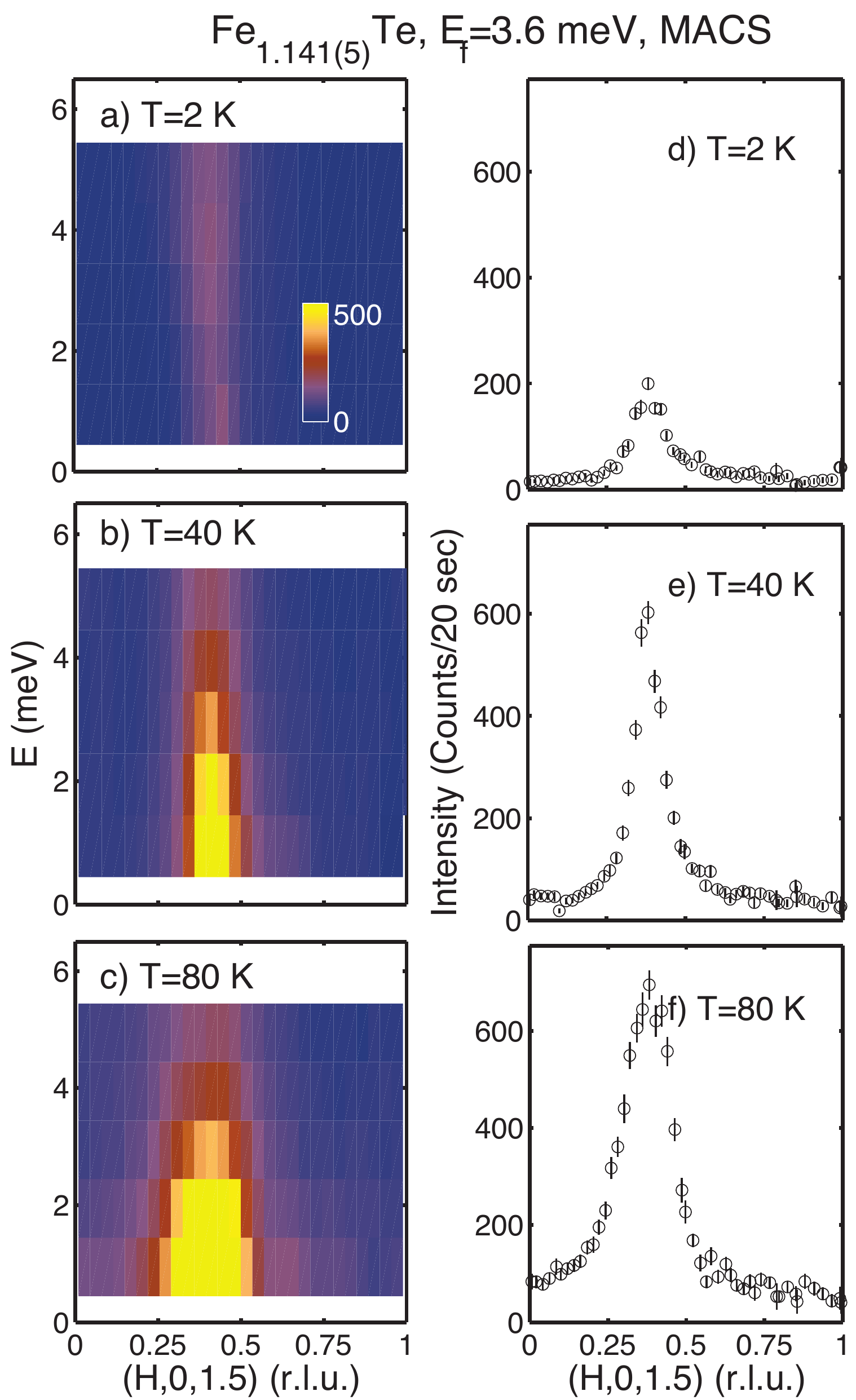}
\caption{\label{constantQ_incom} [color online] Constant-Q scans through the magnetic correlations in Fe$_{1.141(5)}$Te. $a-c)$ show constant Q slices taken over (H,0,L$\pm$ 0.05). $d-f)$ illustrate constant E=2.0 meV cuts integrating around L $\pm$ 0.05.  The magnetic intensity at 80 K is peaked at the incommensurate value of H=0.375 $\pm$ 0.007 $rlu$. }
\end{figure}

Figure \ref{constantQ_incom} shows the temperature dependence of the incommensurate fluctuations in iron rich Fe$_{1.141(5)}$Te.  The magnetic order is characterized by a spiral phase and this is reflected by the gapless excitations in Fig. \ref{constantQ_incom} $a)$ as reported previously.  On warming, the low energy fluctuations become stronger as shown in panels $b)$ and $c)$.  Unlike the commensurate sample discussed above, the structure of the fluctuations in momentum remain unchanged through T$_{N}$ as illustrated by the constant E=2.0 meV scans shown in Fig. \ref{constantQ_incom} $d-f)$.   The persistence of the incommensurate fluctuations below T$_{N}$ and the lack of any strong temperature dependence in the wavevector is different from the commensurate sample discussed above where the magnetic fluctuations are completely gapped at T$_{N}$.

\subsection{Poorly metallic or ``semiconducting" behavior from spin fluctuations and change across $\sim$ 11 \%}

The sudden drop in the resistivity at T$_{N}$ in weakly interstitial iron doped Fe$_{1.057(7)}$Te combined with the abrupt loss of incommensurate spectral weight at low-energies is strongly suggestive that the resistivity is related to a strong coupling to low-energy magnetic fluctuations.   To test whether the gapping of the fluctuations and the strongly sub critical incommensurate fluctuations above T$_{N}$ can account for the semiconducting to metallic transition nearly coincident with T$_{N}$, we have calculated the portion of the resistivity due to spin fluctuations.  We have used the following formula predicted from itinerant magnetism (Ref. \onlinecite{Moriya90:59}),

\begin{eqnarray}
\rho(T) \propto T \int_{-\infty}^{\infty} {E \over T} d\left({E \over T} \right) {{e^{E/T}} \over {(e^{E/T}-1)^{2}}}\int d^{3}q \chi '' (\vec{q},E).
\label{equation_rho} \nonumber
\end{eqnarray}
 
\begin{figure}[t]
\includegraphics[width=8.5cm] {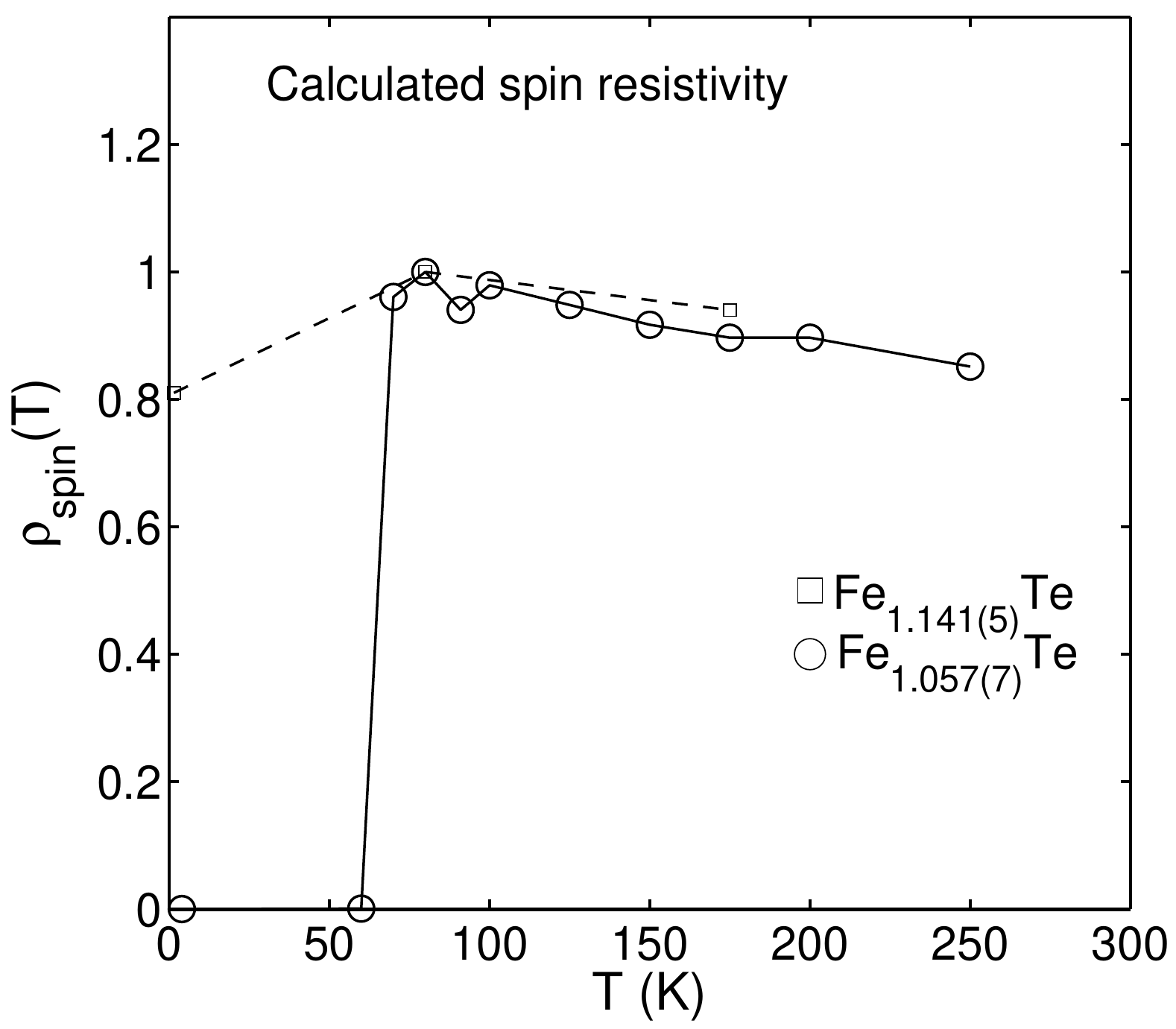}
\caption{\label{spin_rho} The calculated contribution to the resistivity from scattering off spin fluctuations in commensurate and incommensurate samples.  The calculation is described in the main text.}
\end{figure}

\noindent  Here $\chi''$ is the spin susceptibility and is related to the measured intensity via the relation $I(\vec{Q},E)\propto S(\vec{q},E)={1\over \pi} [n(E)+1] \chi''(\vec{q},E)$.  For the calculation, we have taken the temperature dependence of the width in momentum along both [H,0,0] and [0,0,L] from our thermal triple-axis work on PUMA.  We have used the experimental energy dependence extracted from the thermal triple axis PUMA results (Fig. \ref{spin_rho}) for the temperature dependence of the incommensurate fluctuations.  For the energy integral, we have truncated it over the energy range $\pm$ 20 meV by extrapolating our thermal triple-axis results.  While we have not probed the energy gain (negative energy transfer) component in great detail, the combination of detailed balance and the corresponding temperature factor in the integral in Eqn. \ref{equation_rho} ensure that this component is fully determined by the measurement.  The analysis then relies on the approximation that the high energy magnetic spectrum is comparatively temperature independent - an approximation that appears to be substantiated by spallation source experiments.~\cite{Zaliznyak11:107}  A similar calculation has been performed to this for the lightly doped monolayer cuprates (La$_{2-x}$Sr$_{x}$CuO$_{4}$) applying $\omega/T$ scaling in order to explain the linear temperature dependence of the resistivity in those series of compounds.~\cite{Keimer91:67} 

The results of the calculation are presented in Fig. \ref{puma_figure} $e)$.  The resistivity from the spin fluctuations reproduce the semiconducting/insulating nature of the resistivity at high temperatures above T$_{N}$.  The sudden gapping of the magnetic fluctuations also corresponds to the sudden drop in the resistivity at low temperatures and is illustrated by our calculation.  Based on this analysis, we associate the temperature dependence of the resistivity to scattering from incommensurate spin fluctuations with $\vec{Q}$ = ($\sim$0.45, 0, ${1 \over 2}$) at high temperatures.  

The resistivity in heavily doped Fe$_{1.141(5)}$Te is semiconducting at all temperatures which is consistent as the magnetic scattering is described by gapless incommensurate fluctuations even below T$_{N}$.  The calculated results for the resistivity from the spin fluctuations is presented in Fig. \ref{spin_rho} $e)$ and are illustrated by the open squares and the dashed line.  The calculation was performed using the spectrum measured in Fig. \ref{constantQ_incom} and show a nearly constant resistivity with temperature.  While the calculation shows that a nearly constant resistivity is reproduced for this concentration, it does not appear to reproduce the increase in the measured resistivity at low temperatures possibly the result from scattering from defects introduced by the interstitial iron or an additional electronic term.  Regardless of this, the spin fluctuations are therefore strongly coupled to the electronic properties in Fe$_{1+x}$Te with the low-energy spin fluctuations providing a route for scattering electrons and enhancing the resistivity.   Because the temperature dependence of the electronic transport appears to be correlated with scattering from spin fluctuations, we refer to the properties as ``semimetallic" or poorly (bad) metallic.

The analysis connecting the resistivity to the spin fluctuations is in line with the analysis from the specific heat discussed above (Fig. \ref{commens_transport} and Fig. \ref{incommens_transport}).  A plot of the entropy as a function of temperature for small and large interstitial iron concentrations shows localized moment behavior for small interstitial iron concentrations and screened or more itinerant behavior for large interstitial iron.  The analysis of the resistivity from spin fluctuations and the correlation with electrical transport establishes a direct coupling between electronic structure and the spin fluctuations.  This supports the notion of more itinerant response with increased charge doping from interstitial iron.

Low-energy magnetic fluctuations were also found to compete with gapped superconductivity in a study of interstitial iron doped Fe$_{1+x}$Te$_{0.7}$Se$_{0.3}$, therefore suggestive that low-energy magnetic fluctuations compete with superconductivity in the iron telluride system.~\cite{Rodriguez11:2,Stock12:85,Tsyrulin12:14,Bendele10:82}  It is interesting to note that several studies have reported weak and spatially short-range incommensurate magnetic order competing with superconducting in Fe$_{1+x}$Te$_{0.5}$Se$_{0.5}$ materials.~\cite{Xu10:82}  These combined results demonstrate that these low-energy fluctuations strongly scatter electrons therefore enhancing the resistivity.  They are therefore detrimental to superconductivity in these materials.  

While the magnetic spectrum is gapped with a similar low-energy structure to commensurate Fe$_{1+x}$Te, the magnetic spectrum in superconductors is concentrated around $(\pi,\pi)$ type positions and not the $(\pi,0)$ position found in the parent material.~\cite{Qiu09:103,Stock11:84,Stock12:85}  The wavevector seems to be a crucial component of superconductivity in these materials along with the gapping of the magnetic fluctuations.

\subsection{Tricritical-like behavior at $x \approx 11 \%$}

\begin{figure}[t]
\includegraphics[width=0.95\linewidth,angle=0.0]{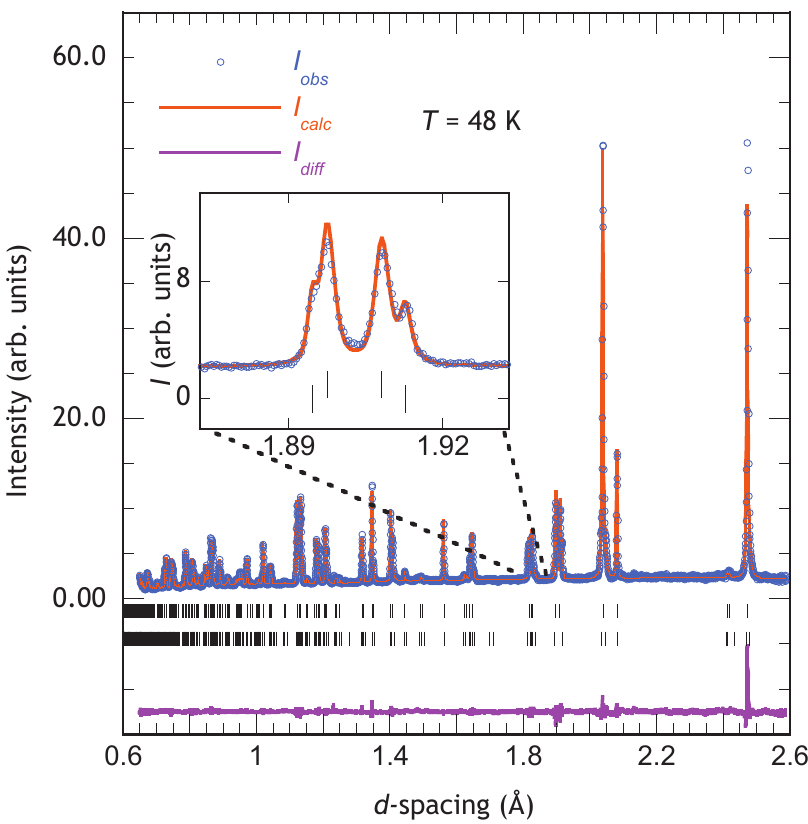}
\caption{[color online]  Observed and calculated neutron powder patterns for Fe$_{1.11}$Te at 48 K from the highest resolution bank of the time-of-flight instrument HRPD (ISIS). The Rietveld refinement converged with an $R_{wp}$ of 10.7 \% and $\chi^2$ of 3.234.  Upper tick marks indicate the Bragg reflections of the orthorhombic $Pmmn$ phase and lower tick marks the monoclinic $P2_1m$ phase.  Inset shows a zoom-in of part of the powder pattern, demonstrating the high-resolution quality of the data to distinguish between the monoclinic and orthorhombic phases even close to the temperature of the nucleation of the monoclinic phase.}
\label{HRPD_figure}
\end{figure}

The thermodynamic and magnetic response for the two extremes of interstitial iron doping outlined above are very different.  The magnetic ordering for small concentrations of interstitial iron is collinear and commensurate while for large concentrations of excess iron, the ordering is clearly incommensurate and spiral.  These properties are also reflected in the magnetic dynamics with commensurate materials showing a gapped excitation spectrum and high interstitial iron concentrations displaying gapless spin fluctuations.  The structural properties are also disparate with small excess iron concentrations showing a monoclinic unit cell while for large it is orthorhombic.  

One of the most striking contrasts between the two extremes in interstitial iron concentrations is displayed in the resistivity as well as the heat capacity.  For small interstitial iron the resistivity displays a sharp ``semimetal" (or poor/bad metallic properties) to metallic transition at the magnetic transition and where the structural distortion also occurs.  For large interstitial iron levels, the transport data shows semi-metallic ( poorly metallic) properties for all temperatures, while showing a peak at around T$_{N}$ where spiral long range magnetic order sets in.  

The transport is qualitatively reproduced by calculating the resistivity from electron spin fluctuations, therefore demonstrating a strong coupling between the electronic and magnetic properties.  These two disparate regions of the phase diagram are separated by a sharp boundary in interstitial iron concentration at $\sim$ 11 \%, illustrated in Fig. \ref{FeTe_phases}.  We now investigate the critical properties near this concentration.  We first discuss the structural and magnetic properties based upon diffraction data obtained from HRPD.  We then show inelastic data from MACS sampling the critical properties and corroborating the conclusions derived from the HRPD.

\begin{figure}[t]
\includegraphics[width=8.6cm] {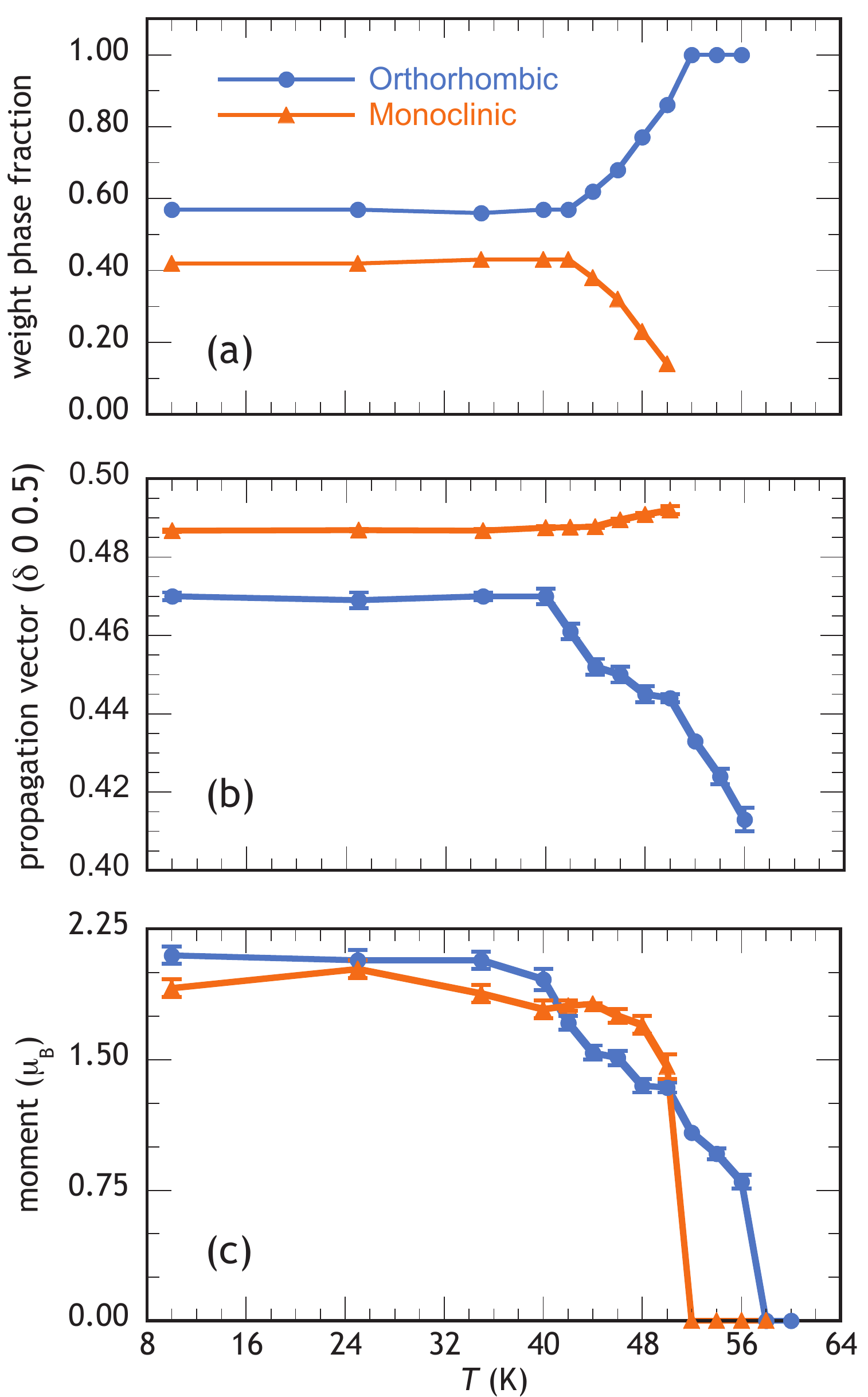}
\caption{\label{1per_phase} [color online] (a) Temperature evolution of the phase fraction in Fe$_{1.11}$Te from neutron powder diffraction data.  The amount of mixed orthorhombic and monoclinic phases is temperature dependent until a percentage of phase fraction locks in at $\approx 40$ K. (b)  The temperature dependence for the incommensurate wavevector, which is strongly temperature dependent for the orthorhombic phase up until the lock-in temperature of 40 K.  (c) The size of the magnetic moment per iron cation as a function of temperature for each of the phases. }
\end{figure}

We have investigated the magnetic and structural properties for a sample with interstitial iron $x$=11$\%$ using the high resolution powder diffraction capabilities available at HRPD (ISIS, UK).  The high resolution allowed us to monitor both monoclinic and orthorhombic phase fractions simultaneously along with the magnetic cross section.  Rietveld refinement with the neutron powder diffraction data obtained from HPRD is illustrated in Fig. \ref{HRPD_figure}.  The scan illustrates the presence of both orthorhombic and monoclinc phases at 48 K represented by a splitting of the nuclear Bragg peaks.  The splitting is well resolved despite the subtle difference between the two diffraction patterns and is based upon similar refinements that details of both phases could be tracked with temperature.

Fig. \ref{1per_phase} shows the properties of both phases in Fe$_{1.11}$Te as a function of temperature.  Panel $a)$ shows the phase fraction of the orthorhombic and monoclinic components separately, and panel $b)$ illustrates the magnetic propagation wave vectors as a function of temperature associated with the two different phases.  The change in the magnetic propagation wavevector with temperature for the orthorhombic phase is suggestive that this high temperature phase transition is likely second order.  Panel $c)$ shows the value of the magnetic moment associated with both phases and again the magnetic moment associated with the orthorhombic phase increases in a continuous manner compared with the monoclinc component again suggestive of this temperature phase transition being second-order.  

Fig. \ref{1per_phase} clearly shows that the Fe$_{1.11}$Te sample is marked by a coexistence between two disparate phases and is suggestive that the transition from commensurate to incommensurate structures as a function of interstitial iron $x$ is first order.   Our conclusions based upon high resolution neutron diffraction and also spanning a variety of concentrations differs to the claim of continuous transitions~\cite{Zaliznyak12:85} or the presence of two phase transitions.~\cite{Rossler_2011}  The phase boundary between these two commensurate and incommensurate phases appears to be very narrow in interstitial iron.  Transport and thermodynamic measurements for similar interstitial iron concentrations previously (Ref. \onlinecite{Rossler_2011}) identified these two transitions.  However, our high resolution neutron studies show that these transitions are associated with two distinct phases near the tricritical point.

\begin{figure}[t]
\includegraphics[width=8.6cm] {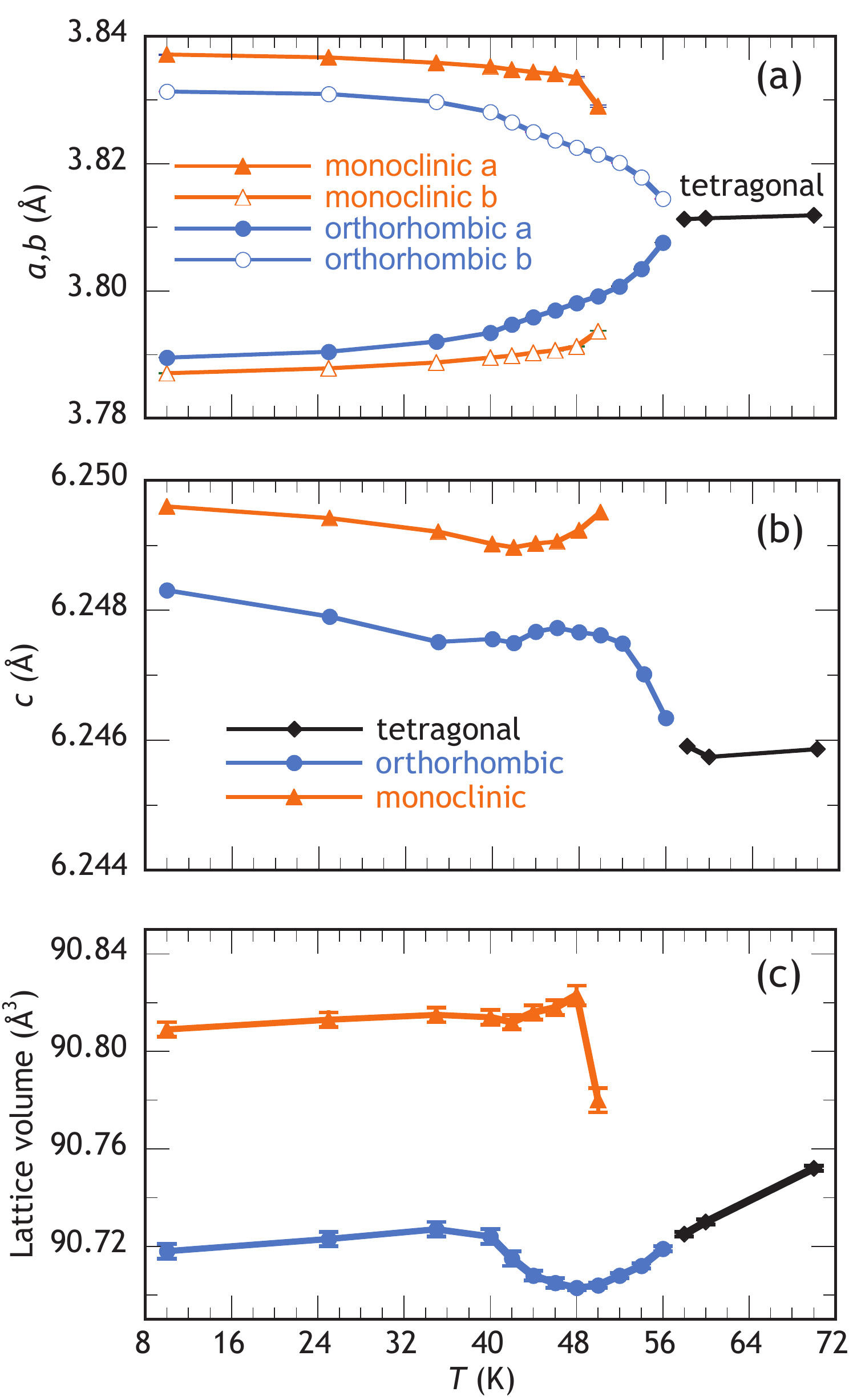}
\caption{\label{1per_lattice} [color online] The lattice constants as a function of temperature for Fe$_{1.11}$Te with $a)$ showing the in-plane $a$ and $b$ values for the two phases. $c)$ shows the c lattice constant and the unit cell volume is shown in $c)$.}
\end{figure}

The lattice constants and unit cell volume are displayed in Fig.\ref{1per_lattice} showing the phase coexistence which exists at all temperatures.  An important point that is highlighted in this plot and seen in Fig. \ref{1per_lattice} $a)$ is that the second order orthorhombic phase sets in before the first order monoclinic.  This is reflected in a plot of the $c$ lattice constant (panel $b)$ and is highlighted by the unit cell volume in panel $c)$.   Based on these data we conclude that with decreasing temperature, the second order orthorhombic phase transition occurs before the first-order monoclinic transition.  Several concentrations near this critical concentration have been studied to construct the overall phase diagram in Fig. \ref{FeTe_phases}.  A fit to the limited order parameter data ($\delta=(a-b)/(a+b)$) near the tetragonal-orthorhombic phase transition in Fig. \ref{1per_lattice} $a)$ yields and exponent of $\beta$=0.36(3). 

\begin{figure}[t]
\includegraphics[width=8.6cm] {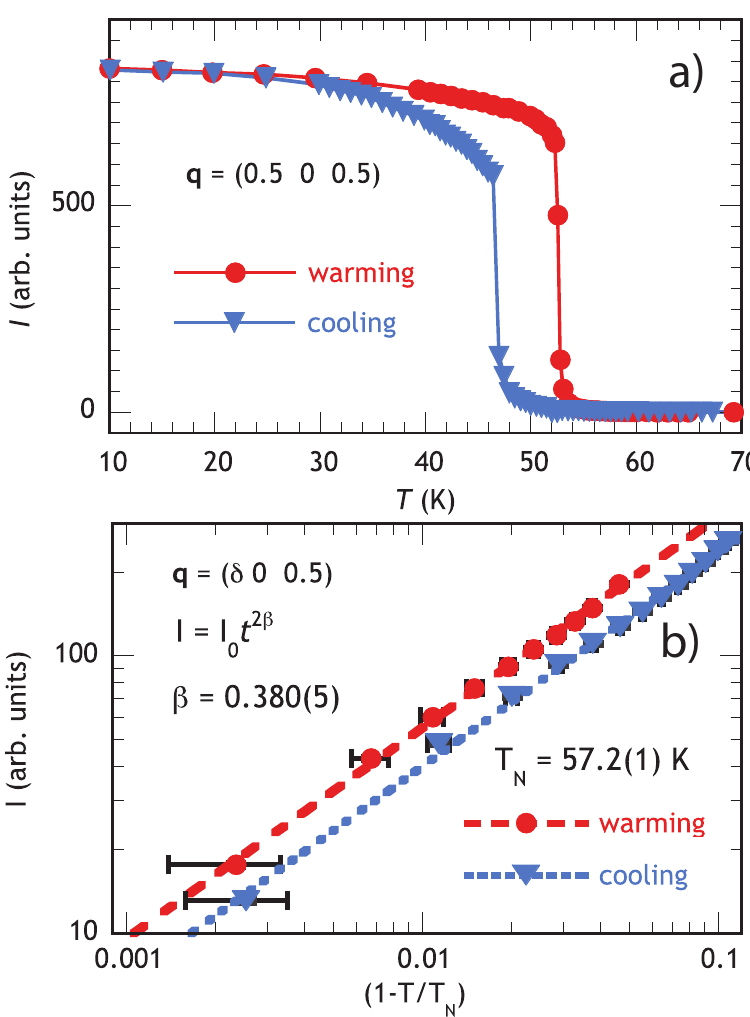}
\caption{\label{1per_order} [color online] The magnetic order parameter plotted for both phases for Fe$_{1.11}$Te for the two different phases.  The data is from the HRPD spectrometer at ISIS.  $a)$ shows the commensurate component and the hysteresis illustrating the first order nature of this transition.  $b)$ shows the magnetic order parameter for the incommensurate phase which displays a second order phase transition.  There is no measurable difference in T$_{N}$ or the slope on both warming and cooling.  The intensity difference reflects the low temperature first order transition.  The exponent $\beta$=0.380(5) belongs to the 3D Heisenberg universality class.}
\end{figure}

Based upon the neutron HRPD data, we plot the magnetic order parameter associated with both phase transition in Fig. \ref{1per_order}.  The magnetic intensity for cooling and warming for the first order commensurate (and monoclinic phase) is shown in panel $a)$ and a fit to the incommensurate second order phase transition (in the orthorhombic phase) is illustrated in $b)$.  Based on panel $b)$ we derive a critical exponent of $\beta$=0.380(5), close to the value of 0.367 predicted for 3D Heisenberg critical properties. The exponent is also in agreement (within error) to the exponent derived for the structural order parameter suggesting a coupling between magnetic and structural order parameters near this critical concentration.  This contrasts with the apparent decoupling of the order parameters for large interstitial iron concentrations in the incommensurate and orthorhombic phase noted above.  It should be noted that there was a measurable difference in the temperature dependence in Fig. \ref{1per_order} $b)$.  No change in T$_{N}$ or the critical exponent was detectable, however there is a difference in the intensity on warming and cooling.  This difference likely reflects the lower temperature first-order transition.

The exponent contrasts with the 2D critical behavior observed for large interstitial iron concentrations reflected from both the order parameter and the magnetic fluctuations.  Therefore, near the tricritical point at $x\sim $11 \%, the critical properties become 3D like before crossing over to 2D behavior at large interstitial iron concentrations.  This occurs while the $c$ lattice parameter decreases with increased interstitial iron concentration.~\cite{Rodriguez_2011b}  

The result for 3D critical properties is somewhat surprising given that both commensurate and incommensurate sides of the phase diagram display clear 2D fluctuations based upon inelastic scattering data presented here and previously.  To investigate this further, we performed neutron inelastic scattering experiments probing the magnetic critical scattering using the MACS spectrometer.  The sample consisted of a 0.5 g single crystal of Fe$_{1.124(5)}$Te previously studied (Ref. \onlinecite{Rodriguez_2011b} using polarized neutrons on SPINS and undergoes magnetic transitions at $\sim$ 50-60 K).  

\begin{figure}[t]
\includegraphics[width=8.8cm] {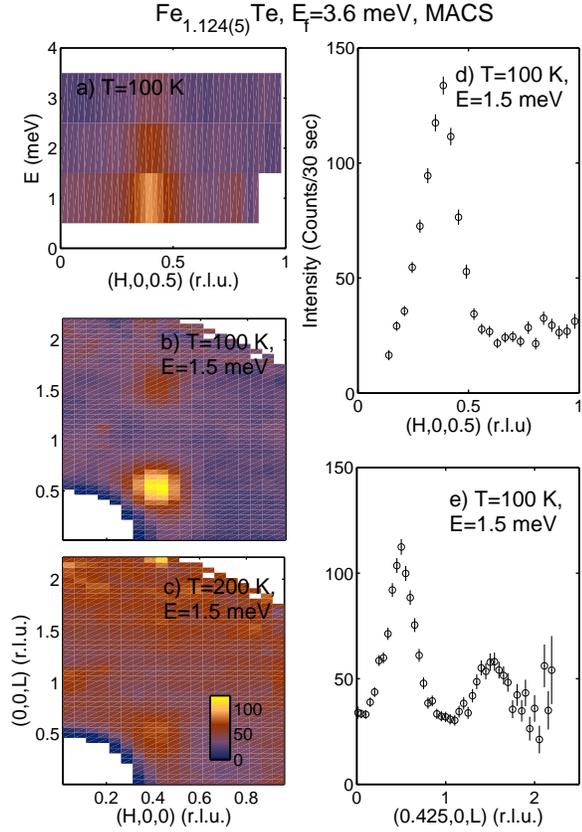}
\caption{\label{1per_inelastic} [color online] A map of the spin fluctuations in the high temperature paramagnetic phase for Fe$_{1.124(5)}$Te.  $a)$ plots a constant-Q slice along $\vec{Q}$=(H,0,0.5$\pm$0.15).  $b)$ and $c)$ show constant E=1.5 meV slices at 100 K and 200 K respectively.  $d)$ and $e)$ illustrate constant energy cuts along $\vec{Q}$=(H,0,0.5$\pm$ 0.15) and $\vec{Q}$=(0.425$\pm$0.075,0,L) respectively at 100 K.  The scattering is peaked at the incommensurate wavevector of H=0.373 $\pm$ 0.008 $rlu$.  The value is within error to that derived for the iron-rich incommensurate sample discussed above.}
\end{figure}

The magnetic fluctuations at high temperatures near this phase boundary is plotted in Fig. \ref{1per_inelastic}.  Panel $a)$ plots a constant-Q slice along the [H, 0, 0.5 $\pm$ 0.15] direction showing that the paramagnetic fluctuations are gapless and indeed incommensurate as expected as a precursor to the orthorhombic incommensurate phase measured by the high resolution powder diffraction experiment described above.  Panel $b)$ and $c)$ show constant E=1.5 meV slices at 100 K and also 200 K.  The plots demonstrate the correlated scattering near ($\sim$ 0.4, 0, 0.5) at 100 K and a decrease in the scattering at 200 K proving the magnetic origin of the correlated scattering.  The scattering is correlated along both H and L directions as demonstrated in panel $d)$ and $e)$.  The width of the peaks along both directions is similar to that observed at high temperatures in Fe$_{1.141(5)}$Te discussed above and were not performed close enough to the critical temperature to observe the difference expected based upon the critical scattering analysis.  The paramagnetic fluctuations which form the basis of the critical response near this tricritical point has strong similarities to the magnetic fluctuations for both commensurate and incommensurate portions of the phase diagram in Fig. \ref{FeTe_phases}.  Despite the different structural, magnetic, and electronic properties at low temperatures, at least in the paramagnetic and tetragonal phase of the Fe$_{1+x}$Te compounds, the spin response is universal being described by gapless incommensurate spin fluctuations where the incommensurate wave vector varies with interstitial iron.

\subsection{Summary and Conclusions}

The phase diagram outlined above as a function of interstitial iron concentration is governed by a region of first order transitions, for small interstitial iron concentrations, which is characterized by a monoclinic ground state and commensurate collinear magnetic order.    Upon increasing the concentration of interstitial iron, this phase gives way to one where the structure is characterized by orthorhombic symmetry and incommensurate spiral magnetic ordering.  This second phase, in contrast, also displays phase transitions at higher temperatures that are best described as second-order.  The two disparate phases are separated by a region of tricritical like beahvior at $x\sim$ 11 \% where phase coexistence is observed through high resolution neutron powder diffraction studies.  

Classically, at a tricritical point an exponent of $\beta$=0.25 should be measured and we have not directly observed this in our series of experiments.  There are a number of possible reasons for this.  Studies on doped Co-doped BaFe$_{2}$As$_{2}$ are suggestive that such classical tricritical properties maybe confined to a very narrow region in doping.~\cite{Pajerowski13:87}  Also, experiments on classic tricritical points in the presence of random fields have shown that the critical exponents maybe strongly altered therefore concealing a clean $\beta$=0.25 exponent.~\cite{Birgeneau82:26}   Furthermore, the lower critical dimensionality near a tricritical point is changed in the presence of random fields making the critical properties and phase transitions likely extremely sensitive to charge doping by interstitial iron.~\cite{Imry79:19,Aharony78:18}    Our assertion of a tricritical point therefore relies on the observation of a line of first order transitions to collinear order giving way to a line of second order transitions for incommensurate order with doping of interstitial iron.  Similar arguments have been applied previously to Mg doped CuGeO$_{3}$.~\cite{Christianson02:66}  Interestingly near the the critical concentration of $x\sim 11\%$, the critical scattering characterized by 3D-type critical fluctuations give way to 2D-behavior for larger interstitial iron concentrations.   Based upon the data in Ref. \onlinecite{Pajerowski13:87}, this is the opposite behavior to what is observed in pnictide compounds where the critical fluctuations cross over to 3D behavior with increased doping towards to the superconducting phase.  At small dopings in the BaFe$_{2}$As$_{2}$ system, a tricritical point exists but the critical response found here in Fe$_{1+x}$Te is quite different to that reported in the 122 compound.

While we have constructed this phase diagram based upon a series of scattering experiments, the phase diagram is also reproduced with high pressure studies on a single concentration.~\cite{Takahashi10:200,Koz12:86}  Pressure appears to have an analogous effect to increased interstitial iron which is consistent with the fact that interstitial iron tends to decrease the FeTe layer spacing.  These results are important as they indicate that the physics and properties we observe is the result of charge doping from interstitial iron and not from purely random field effects accompanying the increased amount of interstitial iron and the concomitant structural disorder.~\cite{Liu11:83}  The coupling between interstitial iron and the electronic properties is reflected in the thermodynamic transport properties and in particular the resistivity measured on the commensurate and incommensurate sides of the phase diagram.  The strong coupling between magnetism and electronic properties is further reflected by high-energy inelastic scattering where the spin excitations are not described by sharp spin-waves, but are broadened considerably and extend to high energies.~\cite{Lumsden10:6}

While our results demonstrate a strong coupling between structural and magnetic order parameters for interstitial iron concentrations less than or near the tricritical point at $x\sim$ 11\%, our interstitial rich samples show evidence for decoupling based on the critical fluctuations of the structural and magnetic orders.  A similar decoupling may occur in pnictide compounds such as CeFeAsO$_{1-x}$F$_{x}$ which shows a divergence between magnetic and structural transitions on increased doping.~\cite{Zhao11:107} Na$_{1-\delta}$FeAs may also illustrate this with a difference between ordering transitions and possibly different critical responses.~\cite{Li_2009}  Therefore, while magnetic and structural order parameters are initially coupled in chalcogenide and pnictide materials, increased charge or structural doping appears to drive the two orders apart.

The tricritical point separates a region of commensurate uniform magnetic order from one that is incommensurate and spatially modulated and therefore is defined as a Lifshitz point.  This point separating collinear and noncollinear regions has been predicted by field theories.~\cite{Xu09:xx}  Given the strong dependence of the magnetic ordering wavevector on interstitial iron concentration and hence charge doping, the magnetic order for large interstitial iron concentration is likely defined by charge dopant induced Fermi surface nesting and this has been reflected in several calculations.~\cite{Han_2009,Singh10:104}  The extended dynamic spin response extends up to very high energies ($\sim$0.5 eV) and seems to be inconsistent with well defined spin-waves from a localized spin structure, further consistent with itinerant magnetism and electronic correlations playing a strong role.~\cite{Lipscombe_2011,Lumsden10:6,Zaliznyak11:107}  Therefore, the effects of interstitial iron doping likely have strong consequences on the Fermi surface topology and hence the electronic properties.    

The results here show critical scattering at positions near the (${1 \over 2}$, 0, ${1 \over 2}$) or ($\pi$,0) positions.  Superconductors based upon anion substitution, of either sulfur or selenium, show magnetic fluctuations near the  ($\pi$,$\pi$) positions at wave vector positions like $\vec{Q}$=(${1 \over 2}$,${1 \over 2}$).   For small concentrations of Se, a competition between ($\pi$, $\pi$) and ($\pi$,0) spin fluctuations have been reported possibly indicating the close proximity of a tricritical point.~\cite{Khasanov09:80,Chi11:84,Wen09:80} The spectrum in the superconductors show a magnetic spectrum that is gapped with a similar value to commensurate ordered Fe$_{1+x}$Te, but located near the  ($\pi$, $\pi$) position.  While the gap in these systems has been interpreted as a resonance mode, magnetic field studies have shown that it does not appear to be strongly correlated with H$_{c2}$ (Ref. \onlinecite{Wang11:83}) unlike classic resonance modes observed in heavy fermion systems (for example CeCoIn$_{5}$ described in Refs. \onlinecite{Stock2008:100,Stock12:109,Raymond12:109,Panarin09:78}).  The issue of spectral weight is also questionable as discussed extensively in the cuprates.~\cite{Stock12:85,Abanov02:89,Kee02:88}  This therefore casts doubt on whether the magnetic gap can be directly related to the superconducting gap as speculated and motivated by work on $d$-wave cuprate and heavy fermion superconductors.

In conclusion, the results of the paper are summarized in Fig. \ref{FeTe_phases}.  The interstitial iron is directly tied to the electronic properties, indicative that charge doping is varying as interstitial iron is tuned.  While low interstitial iron concentrations are characterized by a collinear magnetic order, large institial iron concentrations display noncollinear magnetic order possibly the result of Fermi-surface nesting induced from the charge doping.  While interstitial iron may introduce defects or random fields, the dramatic changes observed show a change in the electronic properties correlated with the interstitial iron concentration.

\subsection{Acknowledgements}

Work at NIST, Oak Ridge, and ISIS was funded by the NRC, Department of Commerce, and by the National Science Foundation under Agreement No. DMR-0944772.  Work performed through the University of Edinburgh was funded by the Carnegie Trust for the Universities of Scotland and the Royal Society of Edinburgh.  We are grateful to N.C. Maliszewskyj for expert technical support for experiments performed on MACS.

%


\end{document}